\documentclass[12pt]{article}
\usepackage{amsmath,amssymb,exscale}
\usepackage{graphicx}
\usepackage{epsfig}
\usepackage{multicol}
\usepackage{color}
\usepackage{hyperref}
\hypersetup{colorlinks,bookmarksopen,bookmarksnumbered,citecolor=verdes,
linkcolor=black,pdfstartview=FitH,urlcolor=rossos}
\usepackage{slashed}

\definecolor{blus}{cmyk}{1,1,0,0.6}
\definecolor{verdes}{cmyk}{0.92,0,0.59,0.4}
\definecolor{rossos}{cmyk}{0,1,1,0.55}

\newcommand{\tmtextbf}[1]{{\bfseries{#1}}}
\newcommand{\tmtextrm}[1]{{\rmfamily{#1}}}

\def\be{\begin{equation}}
\def\ee{\end{equation}}
\def\bea{\begin{eqnarray}}
\def\eea{\end{eqnarray}}

\definecolor{red}{rgb}{1,0,0}

\newcommand{\GL}{{\scriptscriptstyle\rm GL}}
\newcommand{\gappeq}{{\rlap{{\raise}.5ex\text{\ensuremath{>}}}{{\lower}.5ex\text{\ensuremath{\sim}}}}}
\newcommand{\lappeq}{{\rlap{{\raise}.5ex\text{\ensuremath{<}}}{{\lower}.5ex\text{\ensuremath{\sim}}}}}

\newcommand{\I}{\tmtextrm{1{\kern}-.24em l}}
\newcommand{\OO}{{\cal O}}
\newcommand{\NN}{{\cal N}}
\newcommand{\vev}[1]{\langle #1 \rangle}
\newcommand{\wt}{\widetilde}

\begin{document}
\topmargin -1.0cm
\oddsidemargin -0.5cm
\evensidemargin -0.5cm

{\vspace{-10mm}}
\begin{center}
{\vspace{0.cm}} {\Large \tmtextbf{ 
Magnetic Response in the  \\[2mm]
Holographic 
Insulator/Superconductor Transition}} {\vspace{.5cm}}\\
%
{\large  Marc Montull$^a$, Oriol Pujol\`as$^a$, Alberto Salvio$^{a, b}$ and Pedro J. Silva$^{a,c}$}
\vspace{.3cm}

{\it {$^a$ Departament  de F\'isica and IFAE, Universitat Aut{\`o}noma de Barcelona, 08193~Bellaterra,~Barcelona}

\vspace{0cm}

 {$^b$ Scuola Normale Superiore and INFN, Piazza dei Cavalieri 7, 56126 Pisa, Italy}

\vspace{0cm}

{$^c$  Institut de Ci\`encies de l'Espai (CSIC) and Institut
d'Estudis Espacials de Catalunya (IEEC/CSIC),
Universitat Aut{\`o}noma de Barcelona,
08193 Bellaterra, Barcelona}}
\end{center}

\begin{abstract}
We study the magnetic response of holographic superconductors  
exhibiting an insulating `normal' phase.
These materials can be realized as a CFT compactified on a circle,
which is dual to the AdS Soliton geometry.
We study the response under i) magnetic fields
and ii) a Wilson line on the circle.
Magnetic fields lead to formation of vortices and allows one to infer 
that the superconductor is of type II.
The response to a Wilson line is in the form of 
Aharonov-Bohm-like effects. 
These are suppressed in the holographic conductor/superconductor transition 
but, instead, they are unsuppressed for the insulator case.
Holography, thus, predicts that generically insulators display stronger Aharonov-Bohm effects than conductors. 
In the fluid-mechanical limit the AdS Soliton is interpreted as a 
supersolid. Our results imply that supersolids display unsuppressed Aharonov-Bohm (or `Sagnac') effects -- stronger than in superfluids.
\end{abstract}

\tableofcontents
\newpage
\section{Introduction}

Holography has recently been  used to model superconductors (SCs) \cite{Hartnoll:2008vx} with the hope that
it may shed light on the nature of high temperature superconductors, 
which evade the weakly coupled paradigm of BCS theory.  The application of holography to superfluids/superconductors relies on considering a strongly coupled $d$-dimensional conformal field theory (CFT) admitting a gravity dual and in which a global/local U(1) symmetry is spontaneously broken by the vev of some charged operator $\OO$. The gravity dual description is composed of $d+1$-dimensional Anti de Sitter (AdS) gravity with a U(1) gauge field and a charged scalar field $\Psi$ dual to $\OO$ which develops a U(1)-breaking condensate.

In the simplest concrete realization \cite{Hartnoll:2008vx}, the CFT lives in a homogeneous $d-1$-dimensional `plane' at a finite temperature $T$. At high enough temperature, the ground state corresponds to the Reissner-Nordstrom AdS planar Black Brane (or `BB' for short), corresponding to the CFT being in a deconfined plasma state.
At low temperatures a hairy BB with a nonzero smooth profile for $\Psi$ around the horizon is energetically favored and thus the superconducting state arises. 
Since the BB exhibits a finite conductivity, this realizes the superconducting transition in a material which otherwise is a {\em conductor}.

Even more interestingly, the holographic duality successfully overcomes the challenge to describe {\em insulating} materials that display superconductivity at low enough temperatures, like, remarkably, the cuprate high temperature superconductors \cite{Lee:2006zzc}. As first discussed in \cite{Nishioka:2009zj}, the {\em AdS Soliton} \cite{Witten:1998zw} geometry precisely realizes this kind of materials. Indeed, the spectrum of fluctuations over the AdS Soliton has a (roughly temperature-independent) mass gap, signaling an insulating behavior -- which can also be viewed as a confining vacuum state. 
Moreover, the geometry lacks any horizons so heuristically one does not expect any obstruction (other than the gap itself) that these geometries develop hairs. 
Indeed, a large enough U(1) chemical potential $\mu$ eventually overcomes the mass gap associated with  $\Psi$, and it becomes energetically favorable for $\Psi$ to condense near the infrared tip of the geometry.\footnote{The conductor/superconductor transition in the CFT language can be likewise understood as $\mu$ overcoming a certain gap. The main difference is that for the conductor conformal invariance forces this gap to be proportional to the temperature.} 

These encouraging findings motivate us to continue a systematic study of this holographic insulator (the AdS Soliton) and its superconductor phase (the AdS Soliton superconductor or `Soliton SC' for short).
Specifically, in this work we shall study the impact of external magnetic fields on such holographic `materials'.
This will include in particular the response of cylindrical materials under a threading magnetic flux. As an additional motivation, let us point out that recently there have been  discussions  in the condensed matter literature on the response to such a flux, which should undergo a  periodicity change for small enough cylinder radii \cite{Tesanovic,Vakaryuk,Loder,WG,barash}. Our results are perfectly in tune with these conclusions.

The first important fact to notice about the holographic insulator model is that inevitably 
one of the spatial directions of the material must be compactified on a circle -- because the AdS Soliton asymptotics include one compact direction. In fact, this is how the mass gap is generated in the CFT: by compactifying it on a circle with appropriate boundary conditions.

\begin{figure}[t]
  \begin{tabular}{cc}
      {\hspace{-1cm}} 
      \includegraphics[scale=0.35]{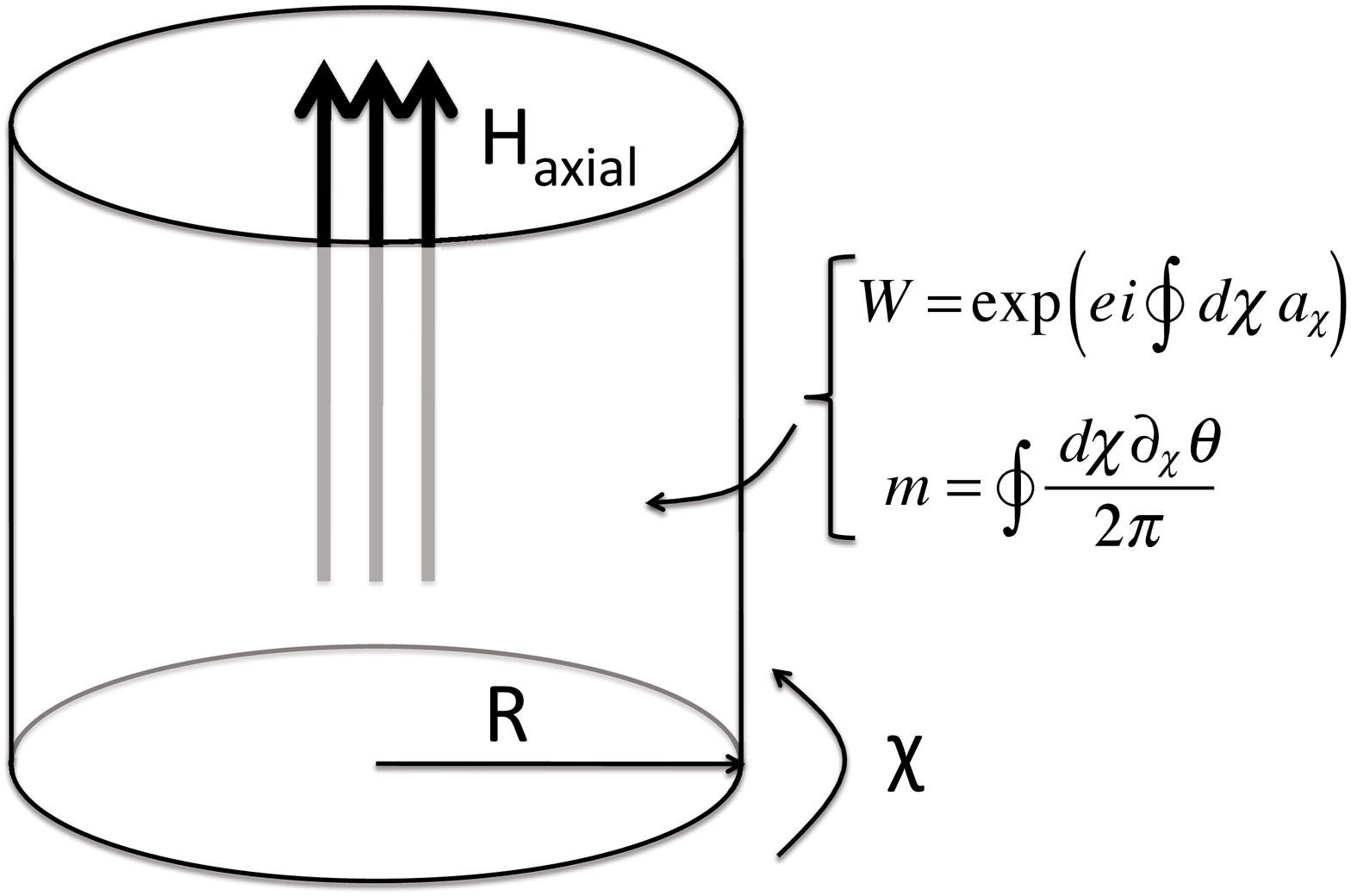}  &
     {\hspace{-2cm}} 
     \includegraphics[scale=0.35]{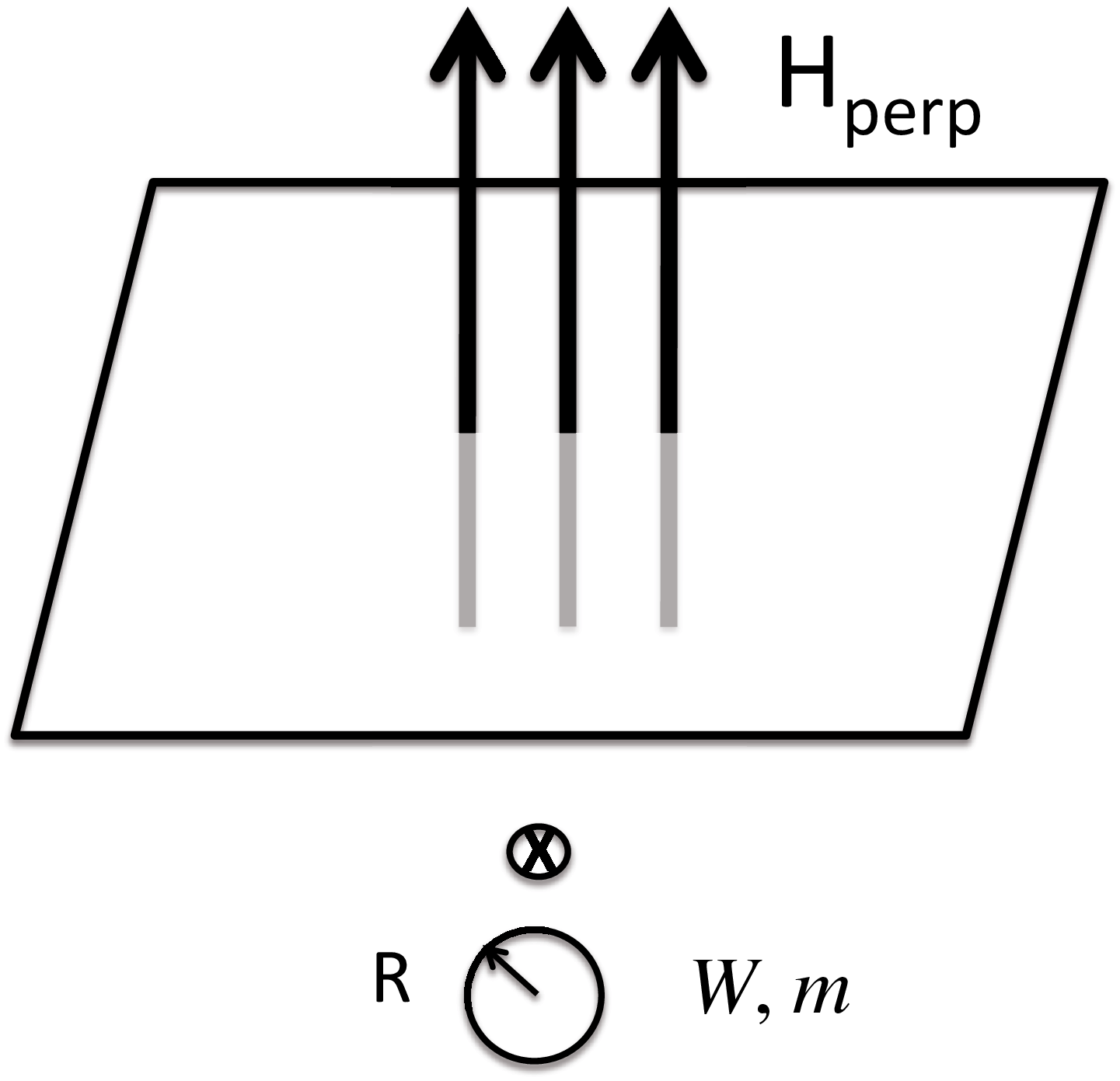}\\
          {\hspace{-2cm}}  (a) 
          &      {\hspace{-2cm}}  (b)
  \end{tabular}
  \caption{{\footnotesize We consider two types of configurations and external fields: 
  (a) a 2-dimensional cylindrical material with nonzero Wilson line $W$. 
  In the lab, this can be accomplished by an `axial' magnetic field threading the interior of the cylinder.
  (b) A planar configuration, in which the cylinder is small and irrelevant for the dynamics.
  In this configuration we shall introduce a magnetic field perpendicular to the plane. 
  We will also discuss the response to both a (perpendicular) magnetic field and a nontrivial Wilson line in this case.}}
\label{pictures}
\end{figure}

This leads us to discuss two distinct situations:
\begin{enumerate}
\item[(a)] Firstly, we will consider 2-dimensional materials physically arranged in a cylindrical geometry of radius $R$, as illustrated in Fig. 1a. In this case  the boundary theory is defined in $2+1$ spacetime dimensions. This will allow us to visualize the dynamics  intrinsically associated with the compact direction more easily. The  `external' gauge field  configuration $a_\mu$ which we will  consider in this case is given by a Wilson line $W$ (see below),
and as we shall show the material responds to it with unsuppressed Aharonov-Bohm (AB) effects. In this article  we shall call `Aharonov-Bohm effects' any flux-dependent effect originating from the fundamental charge carriers (such as the electrons), which may arise in the process of integrating them out. These will be in spirit similar to well known AB-effects that occur in physical materials, such as the existence of persistent currents in small non-superconducting rings (predicted in \cite{PCpred} and observed in \cite{PCobs})\footnote{More on persistent currents in Section \ref{SC-I}.} or the oscillations in the resistivity of certain metals \cite{resistivity}.

\item[(b)] Secondly, we slightly modify this construction in order to describe insulating materials in a planar non-compact configuration. 
This can be accomplished simply by taking the same model with one additional (non-compact) space direction and taking the limit that the compact direction becomes small, $R\to0$. 
For sufficiently small $R$, the dynamics along the circle becomes frozen and the presence of the circle can be ignored in practice for most purposes. Of course, the theory inherits a (large) mass gap and thus the insulating behaviour. 
The superconducting transition still occurs for large enough chemical potentials, $\mu\sim1/R$.
Interestingly, a consistent `compactification' limit for which the superconducting transition persists can be taken in many different ways ({\em e.g.}, 
by sending $R\to0$ with a fixed Wilson line $W$). Therefore really what one describes in this way is a family of insulating materials with slightly different properties (labeled {\em e.g.} by $W$). 
Additionally, once we know how to arrange this type of superconductor in a planar non-compact configuration, it is possible to study the  response to a (perpendicular) magnetic field. 
As we shall see, the superconductor responds to it creating vortices and we can infer that the SC is of type II.

\end{enumerate}

Let us now review the external or control parameters that we will consider. 
The external gauge field that threads the material serves as a control parameter probing the response of the system. 
In presence of the compact direction, there are two types of magnetic fields 
i) an  applied  magnetic field $H_{\rm perp}$ on the non-compact directions, which we will denote also by $H$ henceforth; 
and ii) a gauge-invariant Wilson line\footnote{Here and henceforth we use units of $c=\hbar=1$.}
$$W \equiv \exp\left( e i \oint dx^\mu a_\mu\right)~, $$
where  the integral is done along the compact  direction and $e$ is the charge of the fundamental charge-carriers. 
The physical meaning of $W$ can be seen as follows. In a physical realization in the lab
the electromagnetic field present on the cylinder-shaped material extends to its interior.  
Since, by the Stokes theorem, the circulation of the gauge potential equals the magnetic flux enclosed by the path, 
\begin{equation}\label{axialFlux}
\Phi_{H}^{\rm axial}=\int dS\cdot H_{\rm axial} = {1 \over \,e}\,{\rm arg}(W)~,
\end{equation}
one can think that the Wilson line on the material is generated by the axial magnetic flux. 
For the analysis that we perform in this paper, however, only the quantities defined on the material matter. Therefore it will be completely irrelevant how the Wilson line is `generated', and we shall not make any more reference to $H_{\rm axial}$. 

The response of the system under these control parameters is going to be characterized by i) the possible formation of vortices (in the non-compact directions) which are characterized by the winding number $n$; and ii) by the formation of the so-called {\em fluxoid} configurations in which the  phase $\theta$ of the order parameter winds a number of times around the compact direction. These configurations are characterized by the (gauge-invariant) `fluxoid number' 
$$m \equiv \frac{1}{2\pi} \oint dx^\mu \partial_\mu \theta= \mbox{integer},$$
where  the integral is done along the compact  direction and 
which  plays a role analogue to the winding number associated with  the vortices.  

In Section \ref{Wilson}, to better understand the response to a Wilson line $W$, we first study the system without gauge fields on the non-compact directions, {\em i.e.} with $H=0$. The basic characteristics of this response were  presented in \cite{Montull:2011im}, and here we will further elaborate some points. We argued in \cite{Montull:2011im} that the response to $W$ generically is in the form of Aharonov-Bohm (AB) effects, meaning that the whole effective action acquires an explicit dependence on $W$. 
In particular, this gives a characteristic impact on the phase diagram for the SC transition.
In the limit that the AB effects are absent, the phase diagram displays a periodicity in the direction of the applied axial flux $\Phi_H^{\rm axial}$ given by 
$$\Delta \Phi^{LP} = 2\pi/g_0,$$ 
where $g_0$ is the charge of the condensing operator ${\cal O}$ (for ordinary SCs, $g_0=2e$). We will refer to this as the {\em Little-Parks periodicity} for short.
Instead, the natural periodicity of the AB effects due to the fundamental charge carriers (the `electrons') is  $2\pi/e$, so whenever these quantum effects are important one expects that the LP period will be substituted by the fundamental one, 
$$\Delta \Phi^{fund} = 2\pi/e.$$ 
Equivalently, one can say that in the Little-Parks (LP) regime,
really, the two sub-periods turn out to be degenerate -- and that the degeneracy is uplifted when the AB effects become important.

This degeneracy is simply understood as resulting from the {\em quantum hair} present in superconductors in the form of a discrete gauge charge. 
Indeed, whenever the U(1) gauge group is spontaneously broken by an operator $\OO$ of charge $g_0=Ne$ (where $N$ can be any integer), a discrete gauge subgroup $Z_N$ is realized nontrivially by the `fundamental' fields carrying charge $e$. The discrete charge associated with  $Z_N$ is the simplest form of quantum hair, and is realized in ordinary superconductors with $N=2$.

In these terms, it is clear that there can be a dramatic difference between the Black Brane SC and the Soliton SC,
because (even in AdS) black (horizon-full) classical geometries obviously obey a classical uniqueness theorem, whereas -- as it turns out -- AdS Solitons do not. In turn, this implies that the AB effects are completely suppressed in the BB SC and can be unsuppressed in the Soliton SC.
We will show that they are indeed unsuppressed in the Soliton SC.
 Accordingly, the flux-periodicity exhibited in the insulator/SC transition is $2\pi/e$ whereas  in the conductor/SC it is $2\pi/g_0$.

We are now ready to briefly review the recent developments in the condensed matter literature on the LP effect and its uplifting. Using a microscopic description of the materials \cite{Tesanovic,Vakaryuk,Loder,WG,barash} one concludes that the periodicity in the response to the flux -- or $W$ -- should differ from the LP one for cylinder radii smaller than the (zero-temperature) correlation length $\xi_0$. Heuristically, one can think that the concrete Aharonov-Bohm effect at stake is the one enjoyed by the electrons participating in the pair \cite{WG}. Since $\xi_0$ is the typical pair-size, only for $R<\xi_0$ can an interference take place.
This characteristic behaviour is no different from what we find: 
the length $\xi_0$ turns out to be always smaller than $R$ for the  holographic conductor/SC transition \cite{Montull:2011im}, in which case no AB effect is seen. Instead, $\xi_0$  is always comparable to $R$ for the holographic insulator/SC transition \cite{Montull:2011im}, and the AB effects are noticeable. The only additional feature in the holographic model is that for $R>\xi_0$ in addition to the exponential suppression \cite{Vakaryuk} from the small overlap of the electron's wave function, there is an extra $1/\NN$ suppression. 

Once  the role of $W$ and $m$ in our model is clarified, in Section \ref{vortices} we will study gauge fields on non-compact directions. 
The effect of the magnetic field and vortex solutions in the original HS model of Ref. \cite{Hartnoll:2008vx} has been studied in \cite{Hartnoll:2008kx,Albash:2008eh} and \cite{Albash:2009ix,Albash:2009iq,Montull:2009fe,Keranen:2009re,Domenech:2010nf} respectively. 
Here, in the context of the compactified theory,  we show that such solutions also exist in the broken phase of the superconductor/insulator regime and find them explicitly (the corresponding result in the conductor/superconductor regime can be trivially extracted from previous studies). In order to obtain true superconductor vortices we promote $a_\mu$ to a dynamical gauge
field by making use of the method in Ref. \cite{Domenech:2010nf}.  We also compute  the penetration length of the magnetic field and show that the vortices  are energetically favorable for some window of values of the external magnetic field $H$: $H_{c 1}<H< H_{c2}$; namely these HSs are of type II; interestingly, the high temperature superconductors known so far are of this type.

This paper is organized as follows. In the next section 
we review various aspects of the Holographic Superconductor models which are necessary for our discussion:  their effective field theory (EFT) description in Section \ref{sec:eft}, the physical interpretation of the system when the U(1) gauge field is non-dynamical in Section \ref{sec:SF}, the precise definition of the holographic model in Section \ref{SC-I}
 and the treatment of the gauge field as a dynamical field in Section \ref{dynamical-gauge-field}. 
In Section \ref{Wilson} we describe the response to a Wilson line on the circle. In Section \ref{vortices}, we discuss the response to a magnetic field on the non-compact directions and the formation of vortices. We summarize our conclusions in Section \ref{conclusions}.

Given that we will discuss applications involving different number of spatial dimensions, we will from now on  work in an unspecified number, $d-1$, of spatial (including compact and non-compact) dimensions.

\section{Preliminaries}
\label{sec:background}

\subsection{Effective theories for superconductors and superfluids}
\label{sec:eft}

Before introducing explicitly the holographic model let us now discuss in a model independent way the effect of magnetic fields on  superconductors with compact as well as non-compact dimensions. Our treatment will be similar in spirit to that in  \cite{Weinberg:1986cq}. We are interested in the effective action $\Gamma$ for time-independent configurations of a  U(1) gauge field $a_{\mu}=(a_0,a_i)$, where $i,j=1,...,
d-1$, and  a scalar  field  $\Psi_{\rm cl}$ whose non-zero  value will be responsible for the U(1) breaking. We consider this system at finite temperature $T$ and U(1) chemical potential  $\mu$.  The effective action is a gauge invariant functional of $a_\mu$ and the `order parameter' $\Psi_{\rm cl}$.

In the following we will partly review the model independent results provided in \cite{Domenech:2010nf,Montull:2011im}, which will be used in this paper, and partly extend them in some directions. The reader can therefore refer to \cite{Domenech:2010nf,Montull:2011im} for the proofs of some non-trivial statements given in the following.  Here, like in  \cite{Domenech:2010nf,Montull:2011im}, we  assume that only the magnetic part of the U(1) gauge field is present.

An important observation is that the true  superconductor case corresponds to a dynamical $a_i$ while  a background $a_i$ is suitable to describe  superfluids. The sharpest difference between the two cases arises in inhomogeneous configurations like vortices; we will return to this point when we  explicitly consider such configurations. In this paper we will always use the superconductor language, but all quantities can be translated in the language usually adopted in the superfluid literature: for example a magnetic field corresponds to an external angular velocity performed on the superfluid and arg$(\Psi_{\rm cl})$ to the superfluid velocity potential.

There are two distinct regimes that can be used to define the effective field theory for  $a_\mu$ and $\Psi_{\rm cl}$: when the gradients of the fields are small and when both the fields and their gradients are small. 
In the former case,  the effective action $\Gamma$ can be organized  as a gradient expansion. The latter case reduces to the   so-called Ginzburg-Landau (GL) regime, in which 
the free energy, $F=T\Gamma$, is well approximated by\footnote{
From now on, we will denote  the canonically normalized gauge field as $\hat{a}_i$ and we introduce a rescaled
field $a_i = g_0 \hat{a}_i $.} 
\begin{equation}
  F_{\GL} \equiv \int d^{d - 1} x \left[ \frac{1}{4 g^2_0} \mathcal{F}^2_{ij}+|D_i \Psi_{\GL} |^2 +
  V_{\GL} (| \Psi_{\GL} |) \right]
  \hspace{0.25em}, \label{GL-F}
\end{equation}
where $\mathcal{F}_{ij}=\partial_ia_j-\partial_j a_i$,  $D_i
 \Psi_{\GL}= (\partial_i - ia_i)  \Psi_{\GL}$, $g_0$ is the (renormalized) charge of the order parameter $\Psi_\GL$ and
\begin{equation}
  V_{\GL} \equiv - \frac{1}{2 \xi_{\GL}^2} |
 \Psi_{\GL} |^2 + b_{\GL} |
  \Psi_{\GL} |^4 \hspace{0.25em} .
  \label{GL-P}
\end{equation}
Here we introduce the notation $ \Psi_{\GL}$  for the order parameter to emphasize that it is chosen to have a canonically normalized kinetic term in the GL regime.

We first turn on only the component of the gauge field along the compact direction, $a_\chi$,  where $\chi \in [0,2\pi R).$
A convenient way to analyze this configuration is to perform a decomposition in Fourier modes
(or  ``Kaluza-Klein decomposition") of the  $S^1$ for the theory defined by \eqref{GL-F} in the presence of a Wilson line $W$.  
In order not to introduce further notation, we will from now on denote the non-compact coordinates only by the latin index $i, j\dots$ -- the distinction between when this includes the compact direction will always be clear from the context. 

The Kaluza-Klein (KK) decomposition delivers a tower of massive modes, 
of which we will only keep the most relevant for the dynamics. Thus, truncating  the gauge field to the `zero modes' (with no dependence on $\chi$) for both $a_i$ and $a_\chi$, and to the 
$m$-th fluxoid (or KK mode)  for $\Psi_\GL$, 
$$
\Psi_\GL = \Psi_m\;  e^{{\rm i} m \chi/R}
$$
(here $\Psi_m$  may depend on the non-compact coordinates, and we will take it to be real)
one readily obtains
the following $d-2$ dimensional
{\em tree-level} free-energy in the GL regime:
\begin{equation}
  F_{\GL}^{S_1,\,tree} = 2\pi R \,\int d^{d - 2} x \left[ 
  \frac{\mathcal{F}^2_{ij}}{4 g^2_0} + \frac{(\partial_i a_\chi)^2 }{2 g^2_0}  
   + |D_i \Psi_m |^2 +\left(a_\chi-\frac{m}{R}\right)^2 |\Psi_m|^2+
  V_{\GL} (|\Psi_m|) \right]
  \hspace{0.25em}. \label{GLcomp}
\end{equation}
Let us focus on the GL theory for simplicity: the main conclusions we will find for the set-up described in the introduction in the absence of $H_{\rm perp}$  hold in general.

One observes that the effective mass-term for every fluxoid mode $\Psi_m$  includes a fluxoid-dependent correction, 
\begin{equation}\label{xiLP}
-\xi_{\GL}^{-2} \to  -\xi_{\GL}^{-2}+2\left(a_\chi-{m\over R}\right)^2.
\end{equation}
The Little-Parks effect immediately follows basically from this observation: 
because of this mass-term in the  SC state, 
the fluxoid channel that minimizes the free energy is basically the one minimizing $(a_\chi-m/R)^2$. 
Then, by increasing $W$ ({\em i.e.}, $a_\chi$) there must be transitions between subsequent fluxoids. 
With no corrections to the above {\em tree-level} free energy, one then expects the transitions between different fluxoids to occur at the values $a_\chi=(k+1/2)/R$ with $k$ an integer. Since all fluxoids 
have identical properties except for a shift in $a_\chi$, the whole SC phase transition must display a periodic dependence in $a_\chi$ with period $1/R$. This is indeed the equivalent of the Little-Parks effect, and its associated  flux-periodicity $\Delta\Phi^{(LP)}=2\pi/g_0$. As we showed in \cite{Montull:2011im} and will review in Section \ref{sec:BH}, this is exactly what happens in the holographic conductor/SC transition.

The other important observation regarding 
\eqref{GLcomp}, 
is that the KK reduction for the  gauge field `delivers' a scalar field, $a_\chi$,
that is massless {\em at tree-level}  -- in fact it has no self-potential. 
This field {\em does} appear in the interaction terms with the KK modes $\Psi_m$ (such as the fourth term in \eqref{GLcomp}) via the combination $m-a_\chi$, which originates from local terms involving the covariant derivative $D_\chi \Psi_m$.
This already suggests that a potential for $a_\chi$ can be generated by quantum effects.
In fact, it is easy to see that upon compactification on the circle, higher-dimensional gauge invariance does not forbid a potential for $a_\chi$. Rather, it only requires it (and the whole effective action) to be periodic with period $\Delta \hat a_\chi = 1/(e\,R)$
 where $e$ is the smallest charge in the theory $e=g_0/N$, taken here to be a generic fraction of the charge of the condensate $g_0$. 
Indeed, the Wilson line is gauge-invariant, and so the effective action is perfectly allowed to acquire a dependence on $W$, which on the background considered here  translates on (among others) a periodic potential for $a_\chi$. To be specific, the GL free energy incorporating the  quantum corrections takes generically the same form as \eqref{GLcomp} but with all the coefficients replaced by functions of $W$,
\begin{equation}\label{Wdep}
\xi_\GL^{-2} \to \xi_\GL^{-2}(W)~,\qquad
b_\GL\to b_\GL(W)~, 
\end{equation}
in addition to a possible additive potential for $a_\chi$ only, which is irrelevant for the SC transition.

Physically, the dependence of the effective action on $W$ can be viewed as an AB effect -- an interaction between the charge-carriers and a `magnetic flux' which appears only at quantum level. This dependence was obtained long ago in the context of KK compactifications in \cite{hosotani}, where the one-loop (self-)potential  for the analog of $a_\chi$ was found explicitly to arise as (an Aharonov-Bohm version of) the  Casimir effect. We are not aware of previous literature where the dependence on $W$ of the GL parameters $\xi_\GL$, $b_\GL$  is computed. However it seems granted that, {\em e.g.}, the fundamental charge carrier contribution to the self-energy of $\Psi_\GL$ will generically lead to a $W-$dependent (or `Aharonov-Bohm') correction to the mass $\delta\xi^{-2}(W)$. 
In a weakly-coupled theory, this AB mass term can be estimated as a one-loop effect,
\begin{equation}\label{xiAB}
\delta\xi^{-2}(W) \;\sim\; \ell\; {f(W)\over R^2}~,
\end{equation}
where $\ell$ is a `loop factor' giving a moderate suppression (typically of order $10^{-2}$), $f(W)$ is an $O(1)$ function while the overall $1/R^2$ factor follows from dimensional analysis and by requiring that in the 
de-compactification limit $R\to\infty$ there should be no effect.
 Of course, in  a strongly coupled theory higher loops may significantly modify the factor $\ell$, to possibly $O(1)$ values. 
Notice that while this represents a quantum correction to the effective mass-squared 
$\xi_m^{-2}$, it is parametrically comparable to the classical contribution $\sim(a_\chi-m/R)^2$ arising in \eqref{Wdep} from the KK decomposition -- in fact it is only suppressed compared to it in a weakly coupled theory.\\

This discussion is meant to illustrate that the Wilson line $W$ exhibits a remarkable property: 
its classical (self-)potential vanishes, but at at quantum level a non-zero effective potential is generated.
Put another way, at classical level all values of $W$ are degenerate, while physically inequivalent. 
At quantum level,  this classical degeneracy is uplifted.
This is very reminiscent to what is known as {\em Quantum Hair}: an observable (such as a charge, or a field) that is measurable quantum mechanically but not classically. 

This is not exactly the case for $W$ since we have just seen in \eqref{xiLP} that the correlation length (indeed a classical observable) around any fluxoid  does depend on $a_\chi$ (though partially, since it is always in the combination $a_\chi-m$). So the identification of $W$ is really only valid concerning its self-potential and it would be more appropriate to call $W$ a quasi-quantum hair. However, for the sake of simplicity from now on we will treat it as one more example of quantum hair.\footnote{In hindsight, one realizes that it is gauge invariance and locality of the higher dimensional theory which enforces that $W$ behaves as a quantum hair -- that it has no classical potential.} See below for the actual rigorous form of  quantum hair which is  relevant to the present setup (which is related to the discrete gauge charge present in a superconductor).

In terms of quantum hair,  
the  main result of \cite{Montull:2011im} can be simply stated as follows: 

~\\
{\em the CFT in a deconfined plasma state (dual to the Black Brane) is insensitive  to quantum hair, 
whereas the CFT in the confining vacuum (dual to the Soliton) is sensitive to it.} \\

Applying this to the Wilson line, it translates as the Aharonov-Bohm  effect being suppressed in the deconfined plasma state and unsuppressed in the confining vacuum state. 
When SC occurs (in either the plasma or confining states), then one expects 
no modification from the tree-level picture  
in the deconfined plasma and so  the conductor/SC transition should exhibit the Little-Parks period $\Delta\Phi_B=2\pi/g_0$.
Conversely,  one expects a large deformation from the tree-level picture 
in the confining vacuum and so the insulator/SC transition should manifest the fundamental flux period
$\Delta\Phi_B=2\pi/e$. 
This is, in brief, the main point which Section \ref{Wilson} is going to substantiate. \\

The last issue which we have to mention at this point is that following the same logic that leads us expect a dependence on $W$ in the quantum effective action, one similarly concludes that there should also be a dependence on the
fluxoid number $m$. As observed in \cite{gia}, from the EFT point of view, one may include the following coupling between the fundamental charge carriers and the phase of the condensate $\theta={\rm arg}(\Psi_\GL)$
\begin{equation}\label{e'}
e' \; j^\mu \partial_\mu \theta ~,
\end{equation}
where $j_\mu$ is the conserved current of fundamental charge carriers (the electrons). This coupling is certainly gauge-invariant, and its strength $e'$ represents an independent charge (from the usual electric charge). If this interaction is present, 
then the fluxoid configurations can give rise to  additional Aharonov-Bohm-like effects which translates in the dependence on $m$ of the effective action \cite{Montull:2011im}. 
Let us emphasize here that the fluxoid number $m$ also plays the role of a kind of quantum hair -- just like the Wilson line $W$, it has no classical `potential'.

Finally, let us be more precise about the actual notion of quantum hair in our setup.
Obviously, there are some particular kinds of combined dependence on $W$ and $m$ which do arise classically, namely,  via the local gauge invariant quantity 
$$|D_\chi \Psi_\GL | = |g_0\hat a_\chi-m/R|\, |\Psi_\GL |.$$ 
The actual quantum hair is properly identified as the (gauge-invariant) magnitude on which the most general classical effective action does not depend. Assuming that the classical effective action (even beyond the GL approximation) involves (powers of) 
local gauge-covariant operators such as $D_\chi \Psi_{\rm cl}$, the relevant quantum hair is the combination of $m$ and $W$ {\em not appearing} in  $\hat a_\chi-m/(g_0R)$ (recall that $\hat a_i$ denotes the canonically normalized field).
More explicitly,  splitting the gauge field as 
$$
\hat a_\chi \equiv m' /g_0R + \wt a_\chi
$$ 
with $m'$ an integer and $\wt a_\chi$ the non-integer part of $a_\chi$ {\em modulo} ${1/g_0R}$. 
The classical dependence is through local gauge invariant combination (or `classical hairs')
$\wt a_\chi$  and on $m-m'$. 
Therefore we identify the quantum hair as the possible choices 
$$m=m'=k$$ with $k=0,... N-1$. These are granted to be degenerate classically and they represent a magnetic counterpart of the usual discrete gauge charge: there is an $N$-fold of them and the  winding number $m$ is locked to the  `magnetic flux' $m'$. %

To summarize this part, we conclude that for a given Wilson line $W$ and fluxoid mode $m$, one expects that there will be AB effects that generically can be incorporated by promoting the GL parameters to be {\em generic} functions of $W$ and $m$, 
$$
\xi_\GL \to \xi_\GL(W,m), \qquad  b_\GL \to b_\GL(W,m).
$$
The computation of $\xi_\GL(W,m), \, b_\GL(W,m)$ from first principles is outside the scope of this article. However, it is easy to see \cite{Montull:2011im} that introducing such a dependence allows one to describe the change from the  flux periodicity from the Little-Parks value $2\pi/g_0$ 
to the fundamental one $2\pi/e$.

\begin{table}[top]
\begin{center}
\begin{tabular}{|l|l|l|}
\hline & {\bf superfluids}  & {\bf superconductors}  \\ 
 & & \\ 
   {\it {\color{blus} vortex $a_\phi$-behavior}}& $a_\phi$ is frozen &  $a_\phi\stackrel{large \, r}{\simeq} n+a_1\sqrt{r}e^{-r/\lambda'}$  \\
  \phantom{asd} &\phantom{asd} &\phantom{asd}  \\
 {\it {\color{blus} vortex energy} }& $F_n-F_0\stackrel{large \, r_M}{\sim} n^2 \ln( r_M/\xi_\GL)- nBr_M^2/2$ &  finite as $r_M\rightarrow \infty$  \\  \phantom{asd} &\phantom{asd} &\phantom{asd}  \\
{\it {\color{blus}1st critical field}}& $ H_{c1}\stackrel{large \, r_M}{\simeq}2\ln( r_M/\xi_\GL)/r_M^2\stackrel{r_M \rightarrow \infty}{\rightarrow }0 $  &  $H_{c 1} = g_0^2 (F_1 - F_0)/2 \pi$\\
  \phantom{asd} &\phantom{asd} &\phantom{asd}  \\
{\it {\color{blus} 2nd critical field}} & $H_{c2}=\frac{1}{2\xi_{\rm GL}^2}$ & $H_{c2}=\frac{1}{2\xi_{\rm GL}^2}$ 
\\
  \phantom{asd} &\phantom{asd} &\phantom{asd}  \\ \hline
\end{tabular}
\end{center}\caption{\footnotesize Comparison between superconductors and superfluids.  The penetration length $\lambda'$ is a model-dependent constant, generically different from the  inverse mass of $a_i$, $\lambda$. The quantity $a_1$ is another model dependent constant. $F_n$ is the free energy  per unit of volume $V^{d - 3}$ (of the space orthogonal to the $(r, \phi)$-plane) of a vortex with winding number $n$.   $B=\partial_r a_\phi/r$ is the total magnetic field, while $H$ is the external one, normalized in a way that it coincides with $B$ at $\Psi_{\rm cl}=0$. In the superfluid case we always have $H=B$.  } \label{table}
\end{table}

~\\

Let us now turn to the case where $a_i$ is along the non-compact dimensions and to vortex configurations. Notice that to this purpose we need $d>2+1$.  In this work we will be considering time-independent vortex configurations with  cylindrical symmetry as the main example of our theoretical framework. We define $(r,\phi)$ as the polar coordinates restricted to $0\leq r\leq r_M$, $0\leq \phi<2\pi$ and  take the  Ansatz $a_\phi=a_\phi(r) , \Psi_{\rm cl}=e^{i n\phi}\psi_{\rm cl}(r)$
and the other components of $a_i$ set to zero, where $n$ is an integer. In Table \ref{table} we give the model-independent form of important quantities associated with the vortex configurations ($n\neq 0$) in terms of the GL parameters
in the superconductor case and, for comparison, in the superfluid case. There and henceforth we assume $r_M$ to be much bigger than the vortex core and the radius size of the magnetic tube passing through the vortex. Another model-independent property
of superconductors is the fact that the total magnetic flux through the vortex line equals $2\pi n$ \cite{Weinberg:1986cq}, and therefore is quantized. This property is crucially based on the fact that $a_i$ is dynamical: it does not occur 
in superfluids.

In the `mixed' case with both a  magnetic field and a Wilson line ({\em i.e.} with nonzero $a_i$ both along the compact and non-compact dimensions) it is still true that the coefficients in $\Gamma$ generic have a dependence on $(W,m)$.
It follows that the properties of the  vortex will depend on the quantum hairs $(W,m)$. Therefore, the vortices also provide a probe of the Aharonov-Bohm effect in the compact direction: if the AB effects are (un)suppressed, 
then the periodicity in the axial-flux \eqref{axialFlux} of the votrex properties is ($2\pi / e$) $2\pi / g_0$.

\subsection{Superfluids and supersolids in a rotating cylinder}
\label{sec:SF}

As is well known, there is a well defined mapping between  superconductivity and superfluidity which allows one to translate all the previously described phenomena into superfluid physics. This translation basically amounts to taking the limit where the U(1)  group becomes a global symmetry and the U(1) gauge potential becomes a non-dynamical external field that can be identified as the velocity of the fluid container.
In the following we shall make this parallelism explicit for the cylindrical configuration.

The first thing to notice is that in the global U(1) limit the black brane
BB corresponds to a normal fluid (a plasma), and the hairy-BB phase (with scalar hair) represents that fluid having developed superfluidity.

In the regime where the AdS Soliton dominates over the BB, there will be a similar superfluidity transition, but starting from  the Soliton.  
Now, it is quite clear that the (hair-less) Soliton must correspond to a {\em solid}, just like in the gauged U(1) case it is an insulator\footnote{What we mean by a `solid' is a material exhibiting an energy gap in the  mechanical deformations.
It would be interesting to see whether this property implies any form of underlying spatial order, but we have nothing to say in this regard here.}. Therefore, one concludes that the hairy Soliton phase corresponds to a {\em supersolid} --  a solid material which undergoes superfluidity.  This form of superfluidity was conjectured to be possible a long time ago \cite{AL} and it has been realized experimentally quite recently in solid ${}^4$He \cite{supersolid} (see \cite{son,supersolid_rev} for recent reviews).
From the holographic perspective,  a  solid/{\em supersolid} transition is not only possible but it is as simple as the transition between the (AdS-Reissner-Nosdstrom) Soliton to a hairy version of itself.

The second important thing to note is that in order to have a closer parallel with ordinary superconductivity, we shall focus here on pairing-based superfluids, such as the case of ${}^3$He. In this case it is still true that i) there is a global U(1) symmetry (the number of ${}^3$He atoms), ii) the condensing operator spontaneously breaks this symmetry  and iii) this operator has charge 2 so a $Z_2$ subgroup is unbroken and realized nontrivially by the ${}^3$He atoms not bound in pairs. 

Accordingly to these ingredients, close to the phase transition there must be a  Ginzburg-Landau effective description in terms of a charge 2 (scalar) operator.  In fact, the GL effective lagrangian takes exactly the same form as \eqref{GL-F} with only two changes: 1) we remove the kinetic term for the gauge field, as it is now not dynamical; 2) we rewrite the terms formally grouped in
$|D_\mu \Psi_\GL|^2$ as $ |\partial_\mu \Psi_\GL|^2 +  a_\mu J^\mu + a_\mu a^\mu\;|\Psi_\GL|^2$  with $a_i$ now interpreted as an external velocity field, and $J_\mu=i(\Psi_\GL^*\partial_\mu\Psi_\GL-\Psi_\GL\partial_\mu\Psi_\GL^*)$ is the conserved current. The crossed term $a_\mu J^\mu$ now plays the role of the source term for $J_\mu$. This allows in particular to still treat $a_0\equiv \mu$ as the chemical potential, and the $a_\mu a^\mu$ term then ensure that a constant $\mu$ leads to a mean field of the form $\Psi_\GL\propto e^{\mu i t}$. As usual, the superfluid velocity is $\propto J_i \propto \partial_i \theta$ where $\theta={\rm arg}(\Psi_\GL)$. The formally covariant form of the current $J_i -2 a_i |\Psi_\GL|^2$ is interpreted as the momentum density in the frame of the container.

Specializing the dictionary to the cylindrical setup  considered here, one  identifies the `magnetic flux'
arg$(W)/e$ as the {\em circulation}, $C=\oint {\bf v} \cdot {\bf dx}$, which can also be viewed as the surface integral of the vorticity. For a closed path winding around a cylinder of radius $R$, this is $C = 2\pi R^2w $. Likewise, the `fluxoid' number $m$ simply maps to the  angular momentum $m$ carried coherently by the condensate.

The fluid-mechanical analogue of the LP effect is identified as the {\em Hess-Fairbank} effect \cite{HF} -- the decrease of inertia of a superfluid in a cylindrical tank rotating with angular velocity $w$.
This effect follows simply in the GL description from the quantization of the angular momentum $m$ along the circular direction. Because of this, the superfluid angular velocity is quantized in units of 
\begin{equation}\label{wHF}
\hbar \over 2M R^2
\end{equation}
with  $2M$ the mass of the pair.
Hence, the superfluid cannot keep with the container velocity unless this is precisely a multiple of \eqref{wHF}. Thus, one expects that in the frame comoving with the container, the response of the superfluid will be periodic in $w$ with period \eqref{wHF}. This is the global analogue to the Little-Parks effect. In terms of the circulation, it corresponds to a periodicity of the phase transition in $C$ given by\footnote{
For general superfluids, the periodicity is the inverse of the mass of the condensing object. Thus, 
the Hess-Fairbank experiment used ${}^4$He, and observed a periodicity given by $h/M_4$ with $M_4$ the ${}^4$He mass. 
 }
\begin{equation}\label{CHF}
\Delta C^{HF}= {h\over 2M} .
\end{equation}

Now, following the same logic as for the superconductors, it is clear that for pairing-based superfluids this is not the end of the story. Certainly the circulation $C$ and $m$ equally behave  effectively as quantum hairs and an Aharonov-Bohm-type effect  of the unpaired charge-carriers can introduce a periodicity twice that of \eqref{CHF}. To be more precise, the relevant AB-type effect in this context is the so-called {\em Sagnac} effect (see {\em e.g.} \cite{anandan,SatoPackard}). 
Indeed, generically the quantum effects  of the unpaired carriers give rise to corrections to the GL parameters with periodicity $\Delta C^{fund} = h / M$. In our setup, this is a consequence of the fact that the
`single-${}^3$He atom' Hamiltonian in the lab frame is\footnote{We ignore here  the coupling between unpaired atoms with the condensate (the analog of \eqref{e'}), which seems to be possible in the global case also. } 
obtained by replacing the momentum in the compact direction $p_\chi$ by $p_\chi-MwR$.
Since the spectrum of $p_\chi$ is quantized with equal spacings and the quantum correction to the GL parameters is expressed as a sum over all the modes, it follows that  
the sum must be a
periodic function of $w$ with  period $\hbar/MR^2$, 
implying a period in the circulation 
$$\Delta C^{fund} = {h\over M}~.$$ 

The results obtained in the following sections translated to fluid-mechanical language can be simply stated as {\em the holographic solid/supersolid transition has a periodicity in the circulation  $\Delta C = h/M$ whereas the holographic fluid/superfluid transition has periodicity  $\Delta C = h/2M$}. 

A similar parallel can be driven between the superconducting and superfluid vortices, and we refer the reader to Ref. \cite{Montull:2009fe,Domenech:2010nf} and Section \ref{vortices} for all details in this respect. In the remainder of the paper, we will mostly use superconductor language.

\subsection{The holographic model}
\label{SC-I}

\noindent Let us now define the holographic SC model which we will study. The model was introduced in \cite{Hartnoll:2008vx,Nishioka:2009zj}, and consists of gravity with a negative cosmological constant $\Lambda$  coupled to a U(1) gauge field $A_{\alpha}$ and a charged scalar $\Psi$ in $d + 1$ dimensions ($\alpha, \beta = 0, 1, ..., d$). The action is given by
\begin{equation}
  S = \int d^{d + 1} x \hspace{0.25em} \sqrt{- G} \left\{ \frac{1}{16 \pi G_N}
  \left( \mathcal{R} - \Lambda \right) \hspace{0.25em} + \frac{1}{g^2}
  \mathcal{L} \right\},\quad  \text{with }\quad \mathcal{L} = - \frac{1}{4}
  \mathcal{F}_{\alpha \beta}^2 - \frac{1}{L^2} |D_{\alpha} \Psi |^2,
  \hspace{0.25em} \label{action}
\end{equation}
where $G_N$ is the gravitational Newton constant, the cosmological constant
$\Lambda$ defines the asymptotic AdS radius $L$ via the relation $\Lambda = -
d (d - 1) / L^2$; moreover we introduced $\mathcal{F}_{\alpha \beta} =
\partial_{\alpha} A_{\beta} - \partial_{\beta} A_{\alpha}$ and $D_{\alpha} =
\partial_{\alpha} - iA_{\alpha}$. For simplicity, we have not added any
potential for the scalar.
We are interested in geometries with a compact spatial direction, so  
we use coordinates $(t,z,\chi,y^i)$ where $z$ is the holographic direction (with  the AdS-boundary sitting at $z=0$), the compact direction is parametrized by $0\leq \chi < 2\pi R$  and $y^i$, with $i=1,...,d-2$, are flat non-compact directions. In addition, we will work at finite temperature, corresponding to a compact Euclidean time direction with radius $\beta=1/T$. 

We will work in the limit  $G_N\rightarrow0$
 taken such that  the gravitational effect of
${\cal L}$ can be neglected. 
In this limit,  the relevant background metrics are either the neutral AdS BB
\begin{equation}
  ds_{\rm BB}^2 = \frac{L^2}{z^2} \left[ -f(z) dt^2 + dy^2_{d - 2} +
   d \chi^2 + \frac{dz^2}{ f (z)}   \right] \hspace{0.25em} \hspace{0.25em} \label{AdSBB}
\end{equation}
or the so called AdS soliton
\begin{equation}
  \qquad ds_{\rm Sol}^2 = \frac{L^2}{z^2} \left[ - dt^2 +dy^2_{d - 2} +
  f (z) d \chi^2 \right] + \frac{L^2}{z^2 f (z)} dz^2 \hspace{0.25em} \hspace{0.25em}, \label{AdSso}
\end{equation}
where $f(z) \equiv 1-(z/z_0)^d$, and, for the AdS BB, $z_0=d/(4\pi T)$  and, for the AdS soliton, $z_0=d R/2$. Since we are interested in the theory at finite temperature, we will perform the Euclidean continuation
with compact time $it \in [0, 1 / T)$. The metrics in (\ref{AdSBB}) and (\ref{AdSso}) are energetically favorable for $T>1/2\pi R$ and $T<1/2\pi R$ respectively. Notice that, in the soliton case, the circle parametrized by $\chi$ collapses to a point at $z=z_0$ defining the end of a cigar geometry, while it remains finite in the BB case.

The standard dual CFT interpretation of this setup consists of  a   $d$ dimensional CFT compactified on a circle and at finite temperature. 
The vacuum expectation value (VEV) of operators and the external fields are recovered by studying the asymptotic behavior of the bulk fields near $z=0$, such as the gauge field\footnote{In this paper we always gauge fix  $A_{z}=0$. }: 
\begin{equation}\label{response}
\langle \hat{J}_{\mu}\rangle = \frac{1}{z^{d-3}}  \partial_z A_\mu |_{z=0}. 
\end{equation} 
One then identifies 
$\langle \hat{J}_{\mu}\rangle$ with the VEV of the  U(1) current carried by the CFT 
(we then identify $\langle \hat{J}_{i}\rangle= J_i$) and  
$a_\mu$ as the   gauge field that couples  to it.
If, say, one chooses Dirichlet boundary conditions where $a_\mu$ is fixed, then the dependence of $\langle \hat{J}_{\mu}\rangle$ on $a_\mu$ encodes the response of the CFT to an external gauge field.
Similarly, from the AdS-boundary behavior of the scalar field $\Psi \rightarrow s+\vev{\OO} z^d/d$ 
one identifies $\OO$ as  the condensing operator that breaks the U(1) symmetry (we identify $\vev{\OO}$ with $\Psi_{\rm cl}$), and  
$s$ as its source term.

The very first exercise that illuminates what is going to happen in this holographic setup is to consider the effect of a Wilson line in the absence of any U(1)-breaking operators, that is, with the gauge field in the bulk only. While in the BB solution the gauge field configuration $A_\chi = const$ is a solution, in the Soliton geometry it is not. One can easily foresee (see Section \ref{sec:sol}) that the boundary condition for $A_\chi$ at $z=z_0$ is $A_\chi(z_0)=0$, simply because in this geometry the $\chi$ circle closes to a point. Then, in the absence of condensate the solution for $A_\chi$ has a non-trivial profile\footnote{Corresponding to the decoupling limit of the Magnetic Reissner Nordstrom AdS Soliton, the double Wick rotation of the Electrically charged RN AdS BB.} implying that the response of the CFT to the Wilson line is to generate a {\em persistent current} (see Section \ref{sec:sol} for the numerical coefficients)
$$
\langle \hat{J}_\chi \rangle^{vac} = {2-d\over z_0^{d-2}} a_\chi
$$
of a very similar nature of the  observed ones \cite{PCpred,PCobs}.\footnote{The persistent currents of \cite{PCpred} are in conducting rings, corresponding to $d=2$. All our equations strictly hold for $d>2$. The case $d=2$ requires a separate study because of the vanishing of the Weyl curvature in 3D gravity, so it is unclear to us at present  whether `holographic rings' have $\langle \hat{J}_\chi \rangle^{vac}=0$ or not.}
Indeed, such a nonzero vev of $\hat{J}_\chi$ 
is always present in the Soliton (superconducting or not). In particular, in the CFT picture we see that in the presence of a threading flux the  holographic insulator responds by building up a persistent current which arises purely as a  vacuum polarization effect -- the simplest manifestation of the  Aharonov-Bohm effects  in our setup.\footnote{Incidentally, in the fluid-mechanical analogue, this seems to imply the existence of a persistent flow for {\em solids} (with appropriate coherence properties) subjected to rotation via a Sagnac effect. To the best of our knowledge, we are not aware that this effect is known or even possibly measurable. We leave  this issue for the future. } 
In contrast, in the BB
$\langle \hat{J}_\chi \rangle^{vac} $  is always zero classically. One would expect it nonzero at quantum level in the gravity description, which translates as $\langle \hat{J}_\chi \rangle^{vac} $ being simply $1/\NN$ suppressed in this case.  

In \cite{Hartnoll:2008vx,Nishioka:2009zj} it was found that turning on a constant chemical potential $A_0\mid_{z=0}=\mu$
 introduces a phase transition at a critical value $\mu_c$,
from which the scalar field acquires a VEV.  Such
critical value is given by \cite{Hartnoll:2008vx, Horowitz:2008bn,Nishioka:2009zj,Montull:2011im} 
\begin{equation}
 \quad \,\mu_c \simeq 31.8 \, (20.4) T\hspace{0.25em}, \hspace{1em} \text{for} \hspace{0.25em}
  \hspace{0.25em} \hspace{0.25em} d = 2+1 \,(3+1) \hspace{0.25em} ,  \label{mu-T}
\end{equation}
\begin{equation}
  \mu_c \simeq \frac{1.81 \, (1.70)}{R} \hspace{0.25em}, \hspace{1em} \text{for} \hspace{0.25em}
  \hspace{0.25em} \hspace{0.25em} d = 2+1 \,(3+1) \hspace{0.25em} , \label{mu-R}
\end{equation}
for the AdS BB and  the AdS soliton respectively. Eqs. (\ref{mu-T}) and (\ref{mu-R}) can be inverted to give respectively the critical radius $R_c$ and temperature $T_c$ in terms of a fixed $\mu$. Let us only add here that the estimate for the  (zero-temperature) correlation length $\xi_0$ in both cases is  $\sim 1/\mu$ or  
$\sim R$ for the BB and the Soliton respectively \cite{Montull:2011im} basically as a consequence of conformality. Because of the form of the phase diagram between the Soliton and the BB \cite{Nishioka:2009zj} this implies that $\xi_0$ is biggest for the Soliton and that in the  BB is is always smaller than $R$. 

Finally, let us mention that the standard choice to solve the bulk equations of motion is to fix  Dirichlet boundary conditions at $z=0$. Then all physical quantities can be expressed in terms of fixed $a_{\mu}$ and $s$. For example, the free energy $F[a_\mu,s]$   is  obtained from
the  $d+1$ dimensional AdS Euclidean action $S_E[a_\mu,s]$ evaluated
with all bulk fields on-shell:
$F[a_\mu,s]= T\, S_E[a_\mu,s]$.

\subsection{Dynamical gauge fields in holography}\label{dynamical-gauge-field}

\noindent

Another possible choice to define the dual CFT theory is to change the boundary conditions of the bulk gauge field
from Dirichlet to Neumann, which promotes $a_{\mu}$ to a dynamical field. Including the dynamics of the gauge field is crucial in superconductivity (see Table \ref{table}); for example, it is obviously necessary to observe  the Meissner effect  and the exponential damping of $B$ far away from a vortex core. A complete discussion on how to perform this step in HSs has been provided in Ref. \cite{Domenech:2010nf}. Therefore, here we only summarize the results that will be used in this paper.
The above-mentioned boundary condition of the Neumann type is
\be
\frac{L^{d-3}}{g^2}z^{3-d}{\cal F}_z^{\,\,\, \mu} \Big|_{z=0} +\frac{1}{e_b^2}\partial_\nu {\cal F}^{\nu \mu}\Big|_{z=0}+J^\mu_{ext}=0\, ,
\label{maxwell2}
\ee
where $e_b$ and $J^{\mu}_{ext}$ are additional input parameters, which represent respectively  a bare electric charge and an external current.

In the particular case $d = 2 + 1$, Eq. (\ref{maxwell2}) works even when the bare kinetic term is removed, by taking the limit $e_b / g
\rightarrow \infty$. In
this limit, Eq. (\ref{maxwell2}) leads to a composite gauge field, or in other
words to an emergent gauge field, as it is shown by the fact that its kinetic
term is induced by the AdS bulk dynamics.
For $d = 3 + 1$, the situation is rather different.
The gauge field $a_{\mu}$ is a state of infinite norm with a logarithmic
divergence in the UV (which can be seen using, for example, the Kaluza-Klein
expansion). Hence, if our intention is to keep the gauge field in the theory
we must renormalize it, changing the UV structure of our theory. A way to do
so is to absorb the divergence in the bare kinetic term in Eq.~(\ref{maxwell2}),
i.e., taking  $e_b$ in a way that
\begin{equation}
  \frac{1}{e_b^2} = \frac{1}{g_0^2} + \frac{L}{g^2} \ln z|_{z = 0} +
  finite \, terms \hspace{0.25em} \label{div} .
\end{equation}
In this way one obtains a finite norm state that corresponds to an external
dynamical gauge field coupled to a CFT \cite{ArkaniHamed:2000ds}. This result is also valid for $d > 4$, except that the logarithmic divergence in
(\ref{div}) is replaced by stronger ones.

In this work we are interested in space-times with one compact space
dimension, such as the AdS soliton background in Eq. (\ref{AdSso}). These are
dual to CFTs with one compact dimension, which lead to a $d - 1$ effective
field theory at low energies. The above statement regarding the
emergence of dual gauge fields needs to be accordingly modified if we are
interested in the low energy regime. For $d = 3 + 1$, we expect indeed an
approximately emergent gauge field at low energies. The Kaluza-Klein approach
tells us that the $2 + 1$ dimensional gauge field has an effective electric charge at the normal phase with
leading behavior (when $\beta z_0$ is large) given by
\begin{equation}
  \frac{1}{e_{2+1}^2} \propto R \ln (\beta z_0) \hspace{0.25em},
\label{running}
\end{equation}
where $\beta$ is an energy scale coming from the renormalization. Then, in the
dimensional reduction limit, $R \rightarrow 0$, we can have a finite $e_{2+1}$
even if we remove the bare kinetic term by taking $\beta z_0 \rightarrow \infty$, in agreement with the emergence of a gauge symmetry.

Finally, let us emphasize  that the distinction between Dirichlet and Neumann boundary conditions is pertinent to distinguish between the superfluid and the superconductor regarding the vortices. 
As for the magnetic response along the compact direction, one can of course define the problem with Neumann boundary conditions, which  would treat $a_\chi$ as a dynamical field. Since we are considering homogeneous $a_\chi$-configurations, $a_\chi$ would be simply driven to the minimum of its effective potential in the absence of external sources $J_\chi^{ext}$. One could even work out how the phase transition depends on a nonzero $J_\chi^{ext}$. However, since the issue of main concern is to identify the periodicity with respect to the magnetic flux $\sim a_\chi$, only the discussion holding  $a_\chi$ fixed is relevant -- reducing in practice to Dirichlet boundary conditions. For this reason, in Section \ref{Wilson} we will not refer to the formulation with Neumann boundary conditions, even though one can of course still treat the dynamical case by switching on appropriate sources.

\section{Response to a Wilson line}\label{Wilson}

In this section we introduce  a non-trivial gauge vector potential along the compact direction only. Therefore, on the AdS-boundary we generically have
\begin{equation}
  a_{\chi} = A_{\chi} |_{z = 0} \neq 0 \hspace{0.25em} . \label{a-chi}
\end{equation}
Eq. (\ref{a-chi}) corresponds to the gauge invariant Wilson line $W =\exp\left(i e \int d\chi \, a_{\chi}\right)$ that
in the cigar geometry (\ref{AdSso}) implies a non vanishing magnetic flux
through the $(z, \chi)$-surface. 
The simplest Ansatz to study this problem is the following
\begin{equation}
  \Psi = \psi (z) e^{i m \chi / R}, \hspace{1em} A_0 = A_0 (z),
  \hspace{1em} A_{\chi} = A_{\chi} (z) \hspace{0.25em}. \label{hol-vortex}
\end{equation}
We want to study a system with spontaneous
symmetry breaking of the local U(1) symmetry and with a chemical potential, thus the
boundary conditions for $A_0$ and $\psi$ at $z = 0$ are given by
\begin{equation}
  s = 0,\quad a_0 = \mu \hspace{0.25em}. \label{bcboth}
\end{equation}

\subsection{Response to $W$ in the conductor/SC transition}
\label{sec:BH}	
Let us first consider the AdS BB, which as already stated corresponds to a conducting CFT plasma. 
The  equations of motion for the Ansatz in (\ref{hol-vortex}) are
\begin{eqnarray}
  z^{d - 1} \partial_z \left( \frac{f}{z^{d - 1}} \partial_z \psi \right) +
  \left[\frac{ A_0^2}{f} - (A_{\chi} -  m / R)^2 \right] \psi & = & 0
  \hspace{0.25em}, \nonumber\\
  z^{d - 3} \partial_z \left( \frac{f\partial_z A_{\chi}}{z^{d - 3}} \right) -
  \frac{2 \hspace{0.25em} (A_{\chi} - m / R)}{z^2 } \hspace{0.25em}
  \psi^2 & = & 0 \hspace{0.25em}, \nonumber \\
  z^{d - 3} \partial_z \left( \frac{\partial_z A_0}{z^{d - 3}} \right) -
  \frac{2 \hspace{0.25em} A_0}{z^2 f} \hspace{0.25em} \psi^2 & = & 0
  \hspace{0.25em} .
\label{hol-vortex-bulk}
\end{eqnarray}
The requirement of regularity  on the above set of equations implies at $z = z_0$
\begin{eqnarray}
  \frac{d}{z_0} \partial_z \psi +(A_{\chi} -  m / R)^2 \psi &
  = & 0\hspace{0.25em}, \nonumber\\
 A_0& = & 0, \nonumber\\
   \partial_z A_{\chi} + \frac{2 \hspace{0.25em}}{d\, z_0} (A_{\chi} -  m / R) \psi^2& = & 0 \hspace{0.25em} .\label{hol-vortex-reg-BH}
\end{eqnarray}

\begin{figure}[ht]
   \begin{tabular}{cc}
    {\hspace{-0.5cm}}
    \includegraphics[scale=0.7]{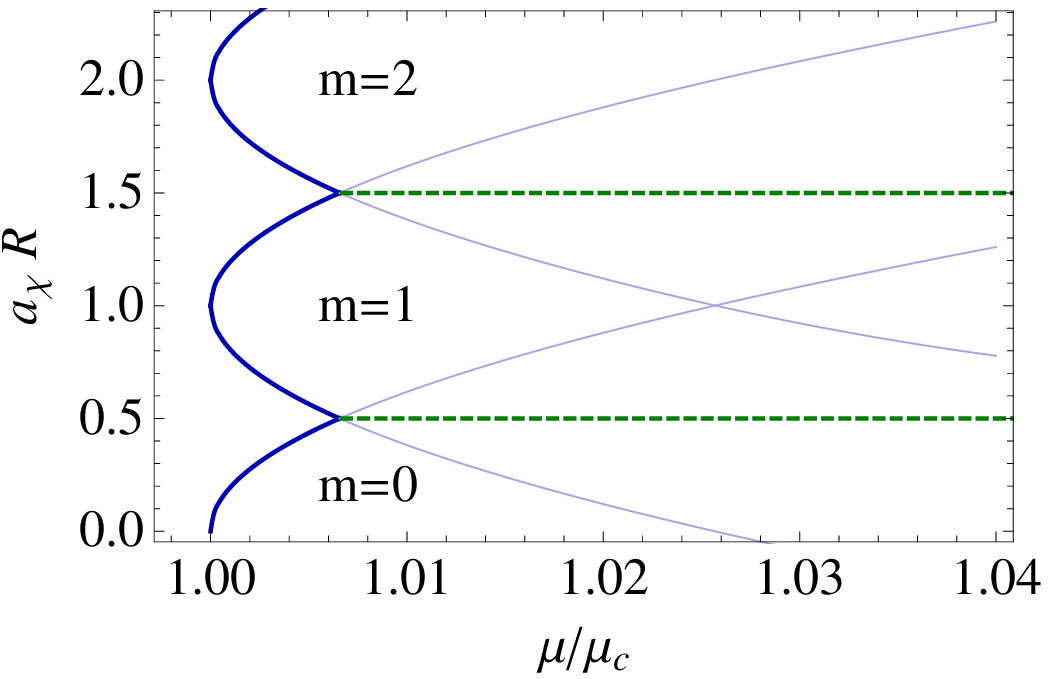}  &
     {\hspace{1.cm}} 
    \includegraphics[scale=0.7]{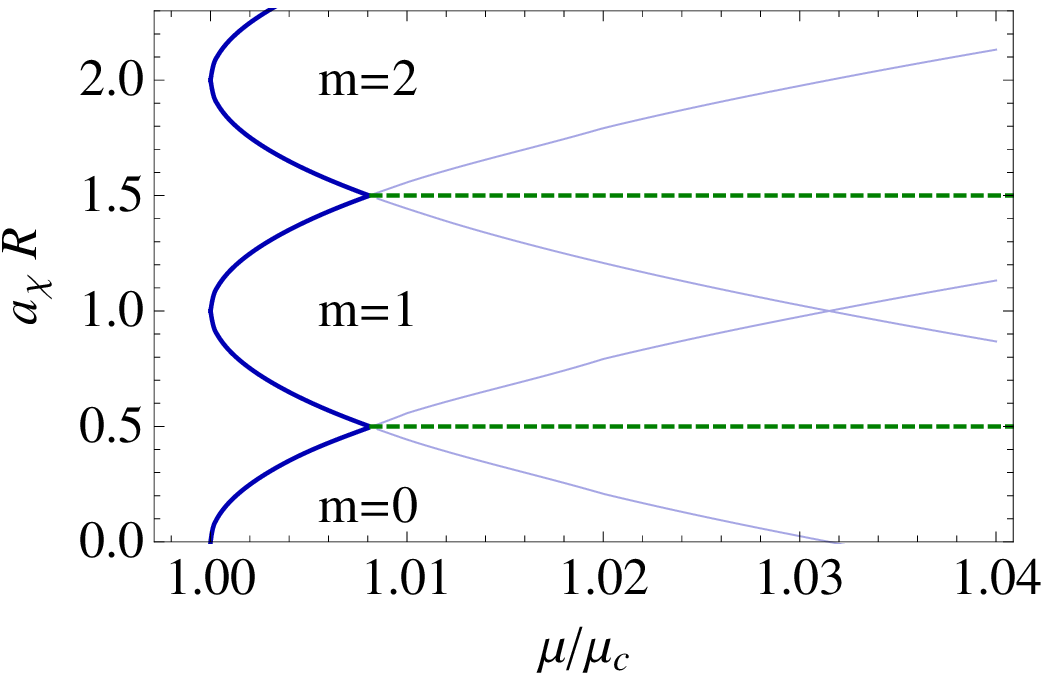}
   \end{tabular}
  \caption{{\footnotesize  Phase diagram for the BB SC at $T = 1/ (\pi R)$. Thick solid blue lines separate the SC and normal phases.  Thin solid blue lines mark the appearance  $m$-fluxoid condensates.
 Dashed green lines separate different fluxoid domains.  On the left $d=2+1$. On the right $d=3+1$.}}
\label{phase-space-BH}
\end{figure}

\begin{figure}[ht]
   \begin{tabular}{cc}
    {\hspace{-0.5cm}}
    \includegraphics[scale=0.7]{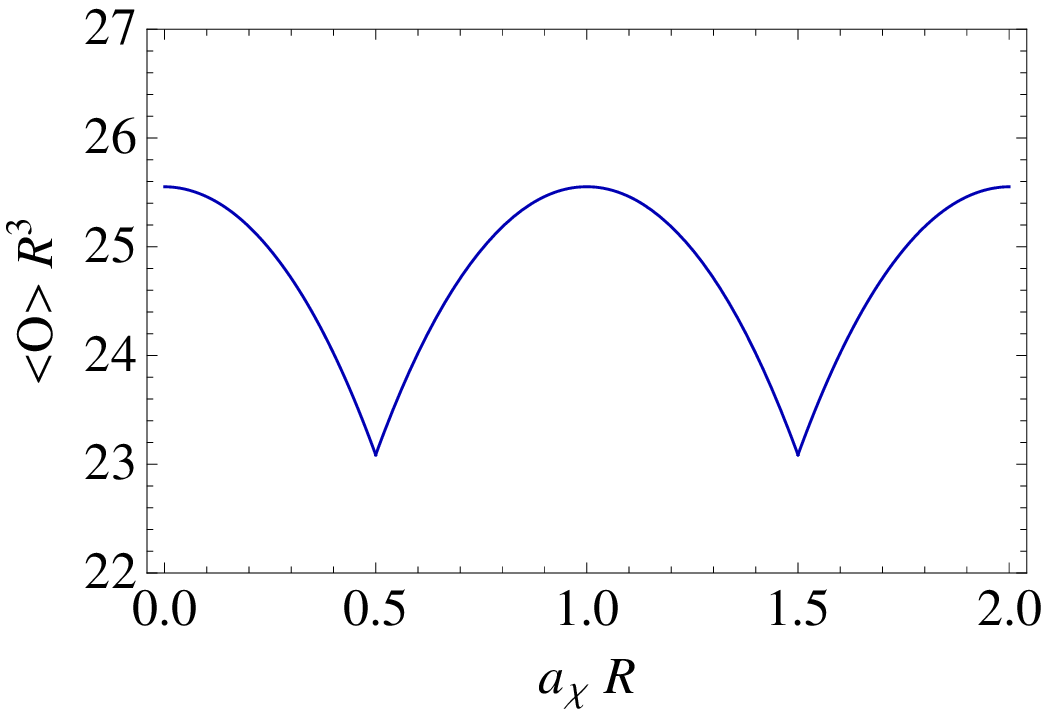}  &
     {\hspace{1.cm}} 
    \includegraphics[scale=0.7]{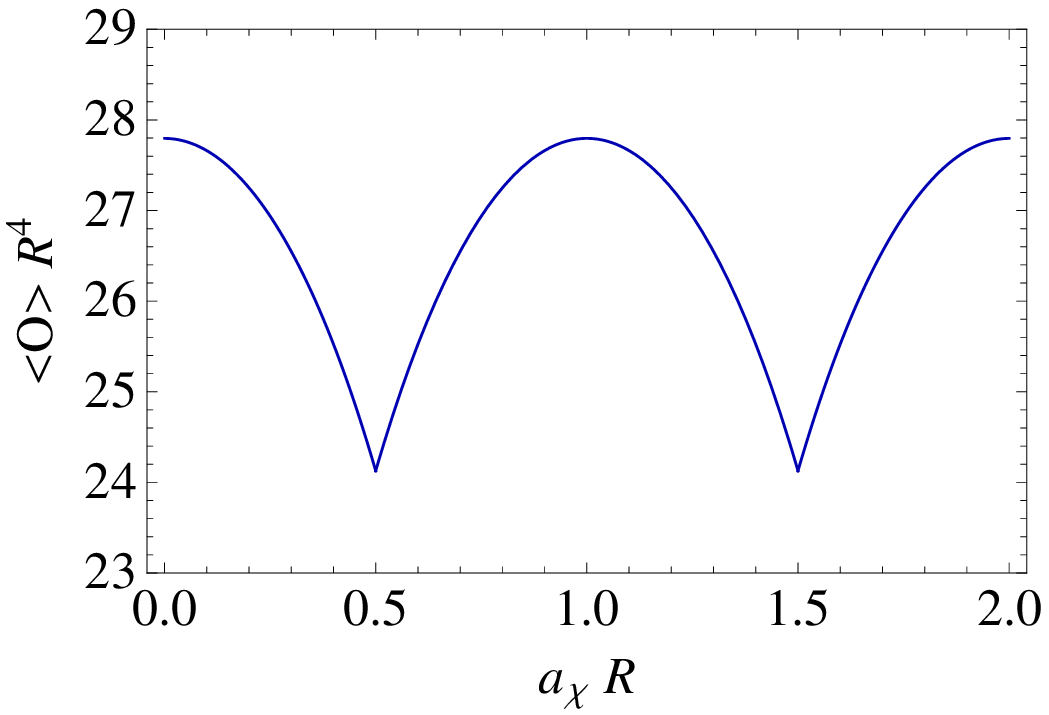}
   \end{tabular}
  \caption{{\footnotesize The modulus of $ \langle{\cal O}\rangle$
as a function of $a_{\chi}$ at $\mu=1.03 \, \mu_c$ and $T=1/(\pi R)$ for the solutions of the form (\ref{hol-vortex}). On the left $d=2+1$. On the right $d=3+1$. }}
\label{OvsachiBB}
\end{figure}

\begin{figure}[ht]
   \begin{tabular}{cc}
    {\hspace{-0.5cm}}
    \includegraphics[scale=0.7]{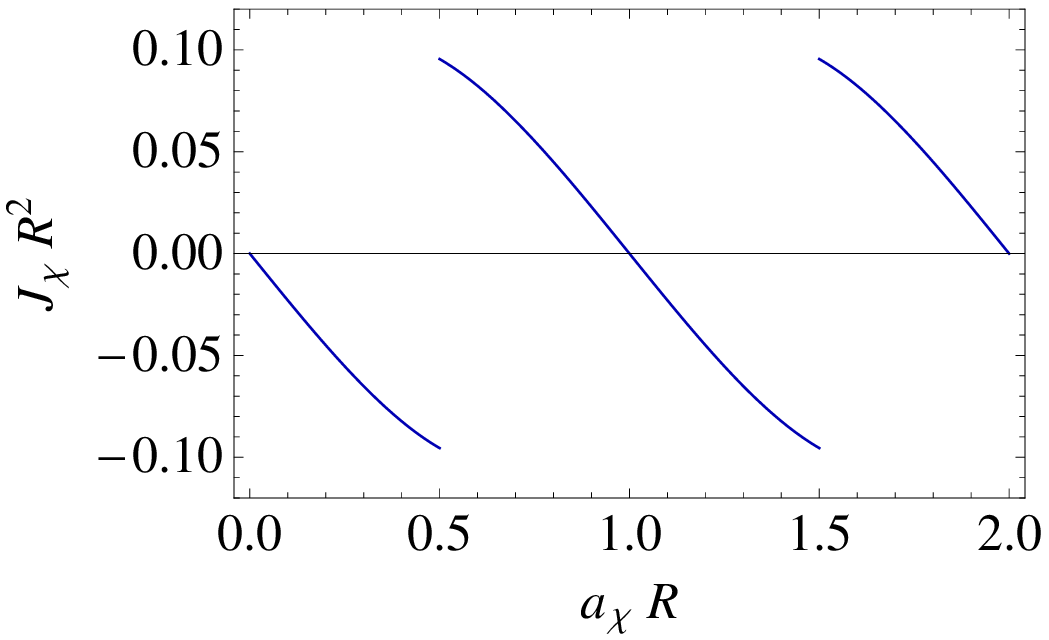}  &
     {\hspace{1.cm}} 
    \includegraphics[scale=0.7]{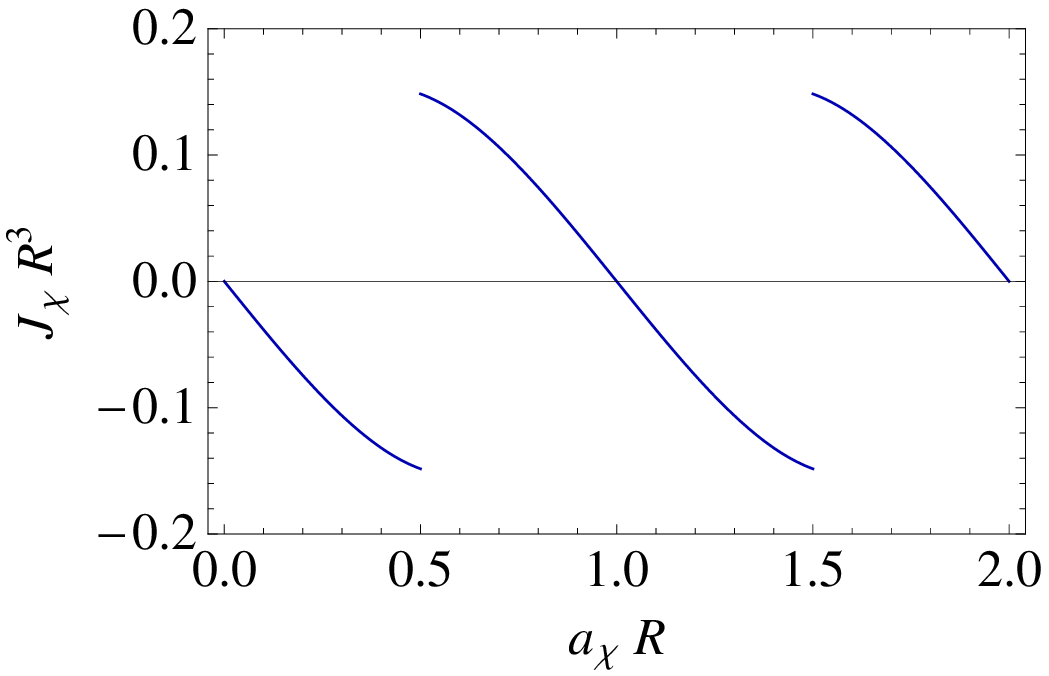}
   \end{tabular}
  \caption{{\footnotesize The current as a 
function of $a_{\chi}$  at $\mu=1.03 \, \mu_c$ and $T=1/(\pi R)$ for the solutions of the form (\ref{hol-vortex}). On the left $d=2+1$. On the right $d=3+1$.}}
\label{JvsachiBB}
\end{figure}

Notice that everywhere in both the bulk equations of motion \eqref{hol-vortex-bulk} and the boundary conditions  \eqref{hol-vortex-reg-BH} $A_\chi$ enters (or can be equivalently written) in the combination $(m/R-A_\chi)$ appearing in the
the local covariant quantity $D_\chi \Psi = i(m/R-A_\chi) \psi  e^{i m \chi / R}$. 
This already suggests that from the CFT point of view the effective action will depend only on local gauge invariant quantities (constructed from the condensing operator), and therefore will display the exact Little-Parks flux periodicity $\Delta \Phi_H=h/g_0$. In other words, the effective action will display no direct dependence on the  `non-local' gauge invariant objects $W$ and $m$ \cite{Montull:2011im}, implying that the Aharonov-Bohm effects are somehow absent in this regime. 

This is confirmed by the form of the phase diagram for the occurrence of superconductivity in the BB, as a function of the magnetic flux $\sim R a_\chi$. 
The phase diagram can be  straightforwardly found by first solving  (\ref{hol-vortex-bulk}) and (\ref{hol-vortex-reg-BH}) for the Ansatz \eqref{hol-vortex} and then finding which of them minimizes the free energy for every choice of temperature $T$, chemical potential $\mu$ and Wilson line $a_\chi$.
The result is shown in Fig.~\ref{phase-space-BH}, 
which displays a periodicity with period 
$$
\Delta a_\chi = 1/R~,
$$
(corresponding to the Little-Parks periodicity in the magnetic flux $\Delta \Phi_H = 2\pi / g_0$) independently of the dimensionality $d$. 
In Figs. \ref{OvsachiBB} and \ref{JvsachiBB} we give instead typical plots of the condensate and the current as a function of $a_\chi$, which display the same periodicity.

From the discussion of Section \ref{sec:eft}, we infer that the Aharonov-Bohm effects are therefore suppressed for the BB, at least when we treat the gravity theory classically. Of course, were we to include quantum effects in the bulk, some dependence on $W$ and $m$ would inevitably appear (with periodicity dictated by the inverse of the charge of the field which is integrated out). 
By the AdS/CFT dictionary, the quantum effects in the bulk translate to subleading effects in the large $\NN$ (number of colors) expansion of the CFT. 
Hence, we realize that rather than absent, the AB effects are simply suppressed at large $\NN$. We proposed in \cite{Montull:2011im} that this can be understood as a consequence of the fact that the limit $\NN\to\infty$ acts as a classical limit \cite{Yaffe}. Therefore, even though in the CFT picture the quantum effects are included, this particular type of effect is sensitive to $\NN$ depending on whether or not the quantum state of the CFT has a classical analogue \cite{Yaffe}. The BB phase corresponds to a deconfined plasma state, which certainly has a classical analogue. Hence in this case the large $\NN$ limit has to render a classical behavior and so the AB effects must vanish in the limit $\NN\to\infty$. 

Instead, the Soliton corresponds to a confining vacuum \cite{Witten:1998zw,Nishioka:2009zj}, which does {\em not} have a classical analogue. Therefore, one expects that the large $\NN$ limit may not result in a classical behavior and therefore the AB effects may still survive for the Soliton. This is indeed what we shall see next. Of course, for the application to the real-world superconductors, we may not have a good candidate parameter that plays the role of $\NN$. However, in the real world case, there is another classical limit -- $\hbar\to0$ -- and one expects that a similar (un)suppression for the two types of vacua can persist.

\subsection{Response to $W$ in the insulator/SC transition}
\label{sec:sol}

Now let us consider the AdS Soliton geometry, which represents the ground state at sufficiently low temperature and which is dual to the CFT in the confining vacuum. 
The  equations of motion for the Ansatz in (\ref{hol-vortex}) take the form
\begin{eqnarray}
  z^{d - 1} \partial_z \left( \frac{f}{z^{d - 1}} \partial_z \psi \right) +
  \left[ A_0^2 - \frac{(A_{\chi} -  m / R)^2}{f} \right] \psi & = & 0
  \hspace{0.25em}, \nonumber\\
  z^{d - 3} \partial_z \left( \frac{\partial_z A_{\chi}}{z^{d - 3}} \right) -
  \frac{2 \hspace{0.25em} (A_{\chi} - m / R)}{z^2 f} \hspace{0.25em}
  \psi^2 & = & 0 \hspace{0.25em}, \label{hol-vortex-eqn}\\
  z^{d - 3} \partial_z \left( \frac{f \partial_z A_0}{z^{d - 3}} \right) -
  \frac{2 \hspace{0.25em} A_0}{z^2} \hspace{0.25em} \psi^2 & = & 0
  \hspace{0.25em} .\nonumber
\end{eqnarray}
This time the requirement of regularity  on the above set of equations implies the following boundary conditions at $z = z_0$
\begin{eqnarray}
  \psi = 0\quad \text{for}\quad m \neq 0\,,\quad - \frac{d}{z_0} \partial_z \psi + A_0^2 \psi &
  = & 0\quad \text{for}\quad m = 0 \hspace{0.25em}, \nonumber\\
  \partial_z A_0 + \frac{2 \hspace{0.25em} A_0}{d\, z_0}
  \hspace{0.25em} \psi^2 & = & 0, \nonumber\\
   A_{\chi} & = & 0 \hspace{0.25em} .\label{hol-vortex-reg}
\end{eqnarray}
The important thing to notice is that the boundary condition \eqref{hol-vortex-reg} now does not depend only on local gauge covariant quantities (such as $D_\chi\Psi$) but it also depends directly on $A_\chi$ at $z_0$, requiring it to vanish. This is of course still a gauge-invariant condition and arises from regularity: since now the spatial circle $\chi$ closes off smoothly at $z=z_0$ ($z$ and $\chi$ represent radial and angular polar coordinates), so the restriction to regular gauge-field configurations automatically demands $A_\chi(z_0)=0$. 
Notice that in our ansatz  \eqref{hol-vortex}   the profile for bulk field  $A_\chi(z)$
is in the homogeneous mode in the $\chi$ direction. Hence, it
coincides up to numerical factors and a logarithm with the (extension into the bulk of the) Wilson line along the $\chi$ direction. Therefore this object is  perfectly gauge-invariant from the CFT perspective. Hence, it is not surprising that the bulk dynamics depends on it. 

More importantly, having the IR condition $A_\chi(z_0)=0$ together with the UV  condition \eqref{a-chi} (demanding a Wilson line in the external gauge field  $a_\chi \sim \log W$)
implies that $A_\chi(z)$ must acquire a nontrivial profile. Hence in the Soliton the presence of a Wilson line leads to the generation of a  magnetic field ${\cal F}_{z\chi}$ in the bulk (localized near the throat of the Soliton). In addition, this nontrivial response of the Soliton to the Wilson line is dual to an Aharonov-Bohm effect from the CFT perspective, since it implies a nonzero conjugate current $\langle \hat{J}_\chi\rangle$ which depends on $a_\chi$.
By the same token the whole effective action acquires some $a_\chi$-dependence. Since the boundary conditions \eqref{hol-vortex-reg} are also directly  sensitive to $m$, one also expects that the effective action acquires dependence on $m$ as well.%
\footnote{
Notice that whereas the bulk description is perfectly local,
the CFT interpretation is in terms of a `non-local' response, understood as a response to the non-local object $W$  (physically, via the Aharonov-Bohm effect discussed in Section \ref{sec:eft}).
}

\begin{figure}[h]
   \begin{tabular}{cc}
    {\hspace{-0.5cm}}
    \includegraphics[scale=0.7]{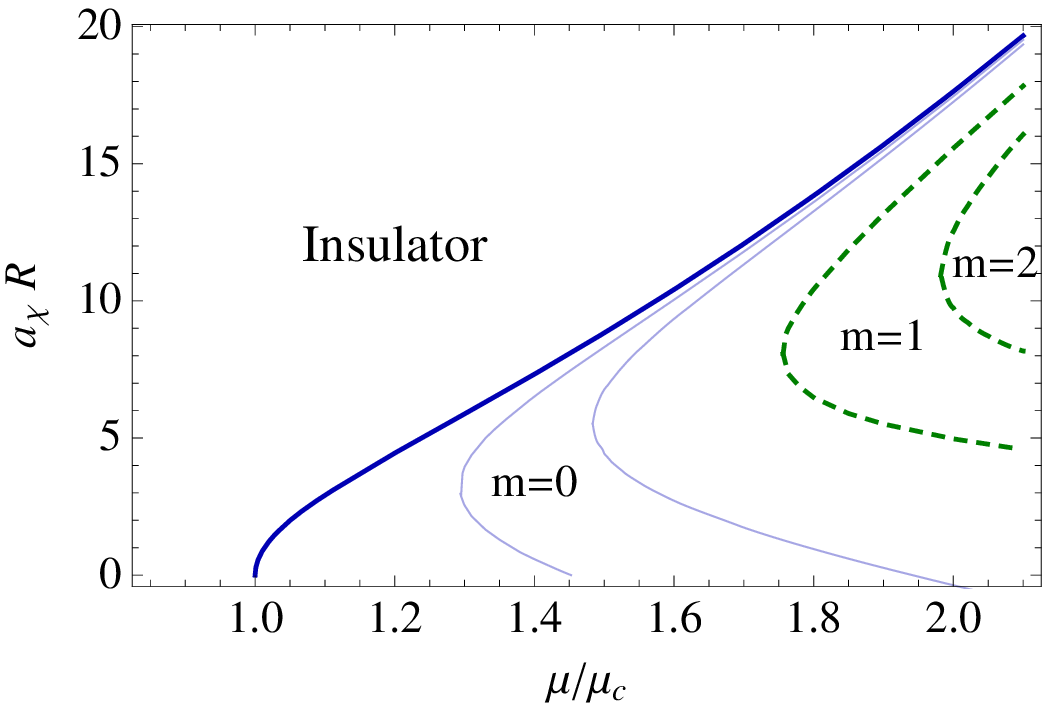}  &
     {\hspace{1.cm}} 
    \includegraphics[scale=0.7]{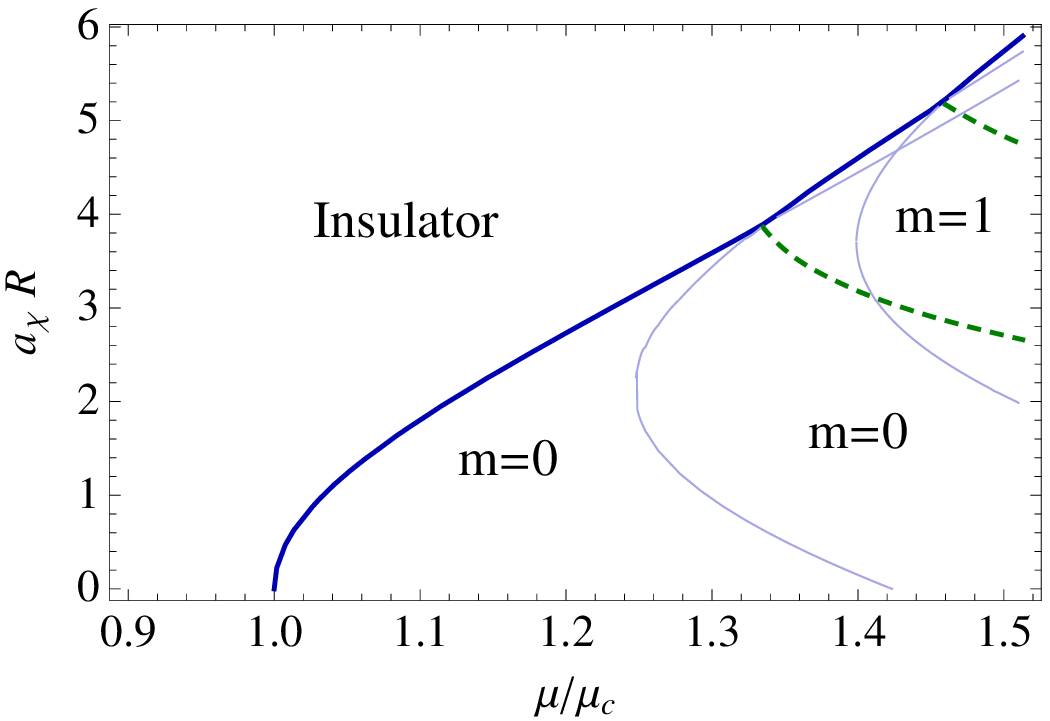}
   \end{tabular}
  \caption{{\footnotesize  Phase diagram for the Soliton. Thick solid blue lines separate the SC and normal phases.  Thin solid blue lines mark the appearance  $m$-fluxoid condensates.
 Dashed green lines separate different fluxoid domains. On the left $d=2+1$. On the right $d=3+1$.}}
\label{phase-space-sol}
\end{figure}

All of this can be checked explicitly, by working out the phase diagram for superconductivity in the Soliton, and we will describe the result of this exercise in the next paragraph. Before that, though, let us work out the response of the CFT in the Soliton phase  without the U(1)-breaking condensate. Assuming for the moment that  $\psi=0$, it is easy to see that the regularity condition $A_\chi(z_0)=0$ imposes that $\langle \hat{J}_\chi \rangle\neq0$. In the decoupling limit, with $\psi=0$, the equations of motion with boundary conditions \eqref{hol-vortex-reg} can be straightforwardly integrated to give $A_\chi(z) = a_\chi (1- (z/z_0)^{d-2})$. Hence, one identifies the CFT response to the Wilson line as
\be 
\langle \hat{J}_\chi \rangle^{vac} = {2-d\over z_0^{d-2}} \;a_\chi~. \nonumber 
\ee 
Let us emphasize that such a response should be interpreted as a normal-phase persistent-current, since it occurs even in the absence of superconductivity -- a kind of vacuum polarization by the Wilson line. 
In the presence of superconductivity, there is of course an additional contribution to $\langle \hat{J}_\chi \rangle$ due to the U(1)-breaking condensate, for which we shall reserve the notation $\langle \hat{J}_\chi \rangle^\OO$ since it is entirely due to the order parameter $\OO$. For comparison, close to the  Ginzburg-Landau regime, $\langle \hat{J}_\chi \rangle^\OO$ should take the form $\sim(m/R-a_\chi)| \Psi_\GL|^2$ with some additional $a_\chi$-dependence hidden inside $| \Psi_\GL|^2$. Since $J_\chi$ is the operator conjugate to $a_\chi$,  a non-trivial response implies that the energetics of different $a_\chi$ configurations are affected. 
However, what matters for the superconductivity transition is only $\langle \hat{J}_\chi \rangle^\OO$ (the contribution from $\langle \hat{J}_\chi \rangle^{vac}$ always cancels out in energy differences). Hence, in the plots below we only display the superconducting contribution to the current, $\langle \hat{J}_\chi \rangle^\OO = \langle \hat{J}_\chi \rangle-\langle \hat{J}_\chi \rangle^{vac}$.

With this in mind, we can straightforwardly obtain the phase diagram for superconductivity in the Soliton including the magnetic flux $\sim a_\chi$ and all the possible fluxoid configurations.   
We have solved Eqs. (\ref{hol-vortex-eqn}) with 
boundary conditions (\ref{a-chi}), (\ref{bcboth}) and (\ref{hol-vortex-reg}). We have found 
that indeed, there exist solutions with $m\neq 0$ that turn on a VEV for the 
scalar operator. In fig. \ref{phase-space-sol} we give the phase diagram varying $a_{\chi}$ and $\mu/\mu_c$. There is a minimal value of $\mu/\mu_c\geq 1$, that depends on $m$, below which 
there is no solution for any $a_{\chi}$. 
Also we have found a region where the solutions with $m\neq 0$ are energetically favorable.
Also, as we get deeper into the region of allowed $m\neq 0$, higher winding
solutions become energetically favorable. Unfortunately, we could not check if these 
solutions are less favorable than multi-centered solutions with the same winding, due to 
the assumed axial symmetry of our Ansatz. 
Nevertheless our results show that solutions with non-trivial winding are energetically favorable 
in the region indicated in Fig. \ref{phase-space-sol}.

\begin{figure}[ht]
   \begin{tabular}{cc}
    {\hspace{-0.5cm}}
    \includegraphics[scale=0.7]{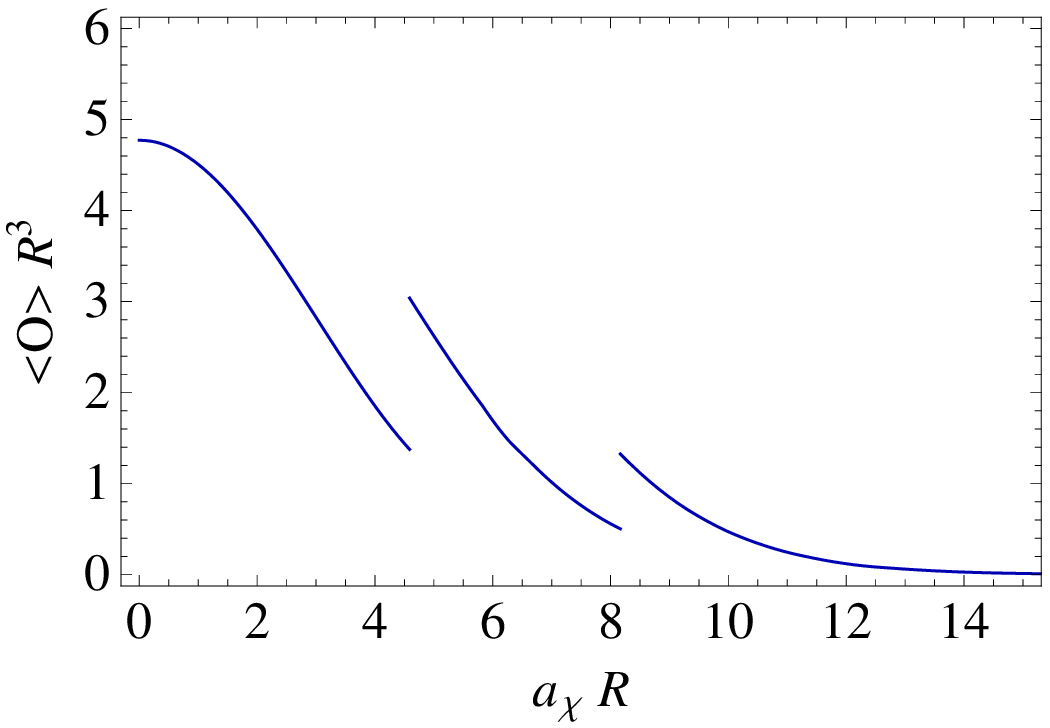}  &
     {\hspace{1.cm}} 
    \includegraphics[scale=0.7]{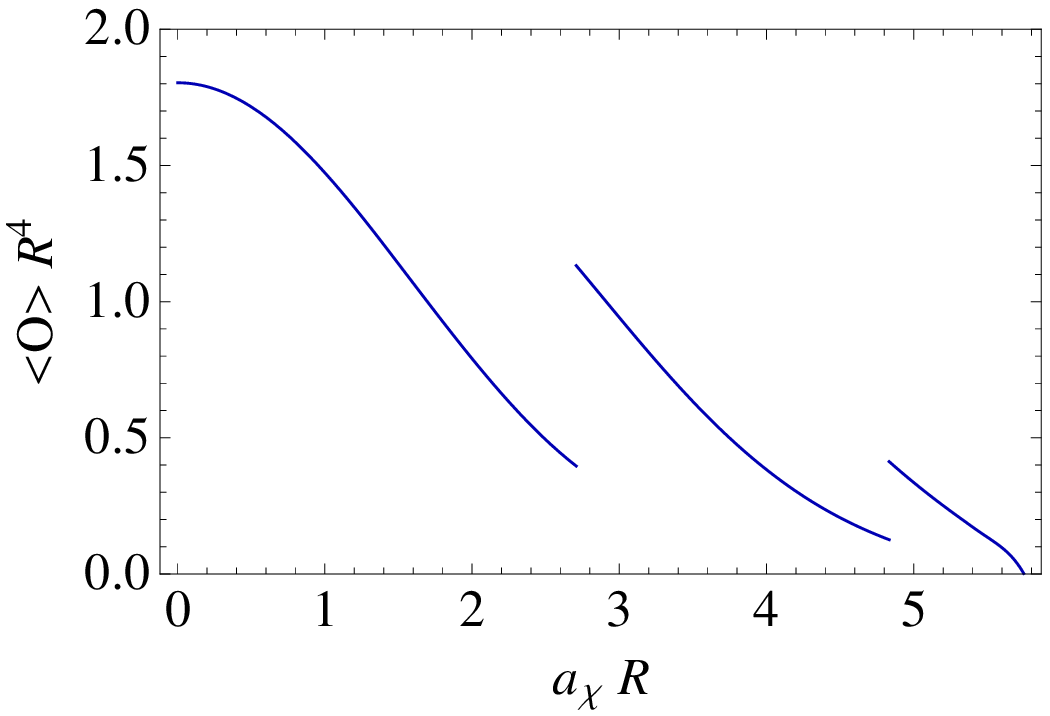}
   \end{tabular}
  \caption{{\footnotesize The modulus of $ \langle{\cal O}\rangle$
as a function of $a_{\chi}$ for the solutions of the form (\ref{hol-vortex}). On the left $d=2+1$ with $\mu/ \mu_c=2.1$. On the right $d=3+1$ with $\mu /\mu_c = 1.5$. 
We observe a jump when the $m=1$ solution becomes energetically favorable at $a_{\chi} \simeq 4.6/R \, (2.7 /R)$for $d=2+1 \, (3+1)$. The second 
jump occurs when the $m=2$ solution becomes the ground state.}}
\label{Ovsachi}
\end{figure}

\begin{figure}[ht]
   \begin{tabular}{cc}
    {\hspace{-0.5cm}}
    \includegraphics[scale=0.7]{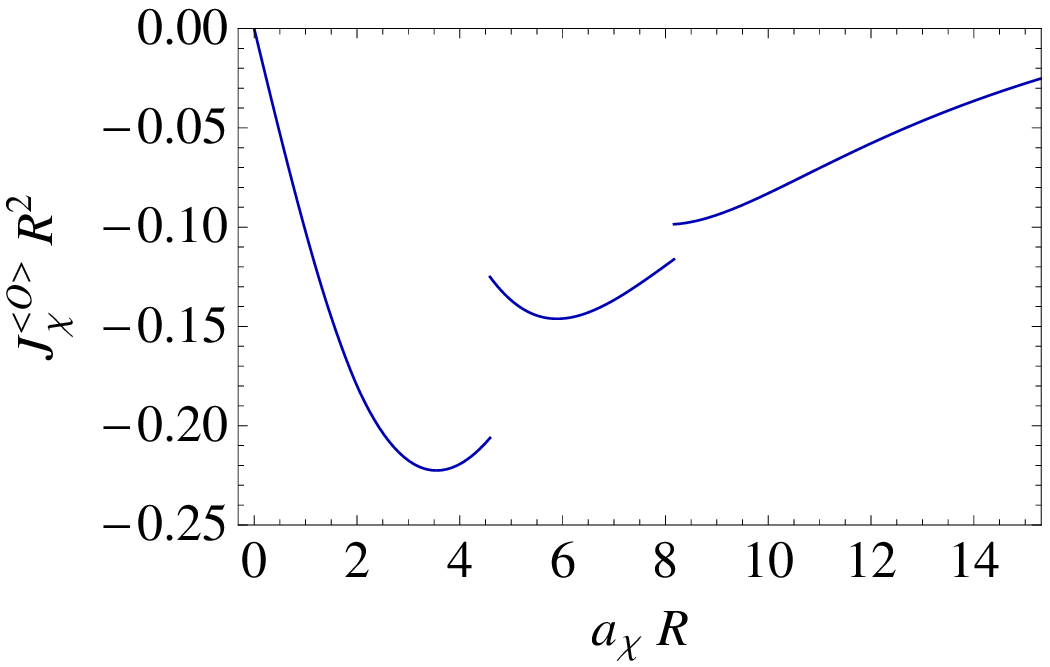}  &
     {\hspace{1.cm}} 
    \includegraphics[scale=0.7]{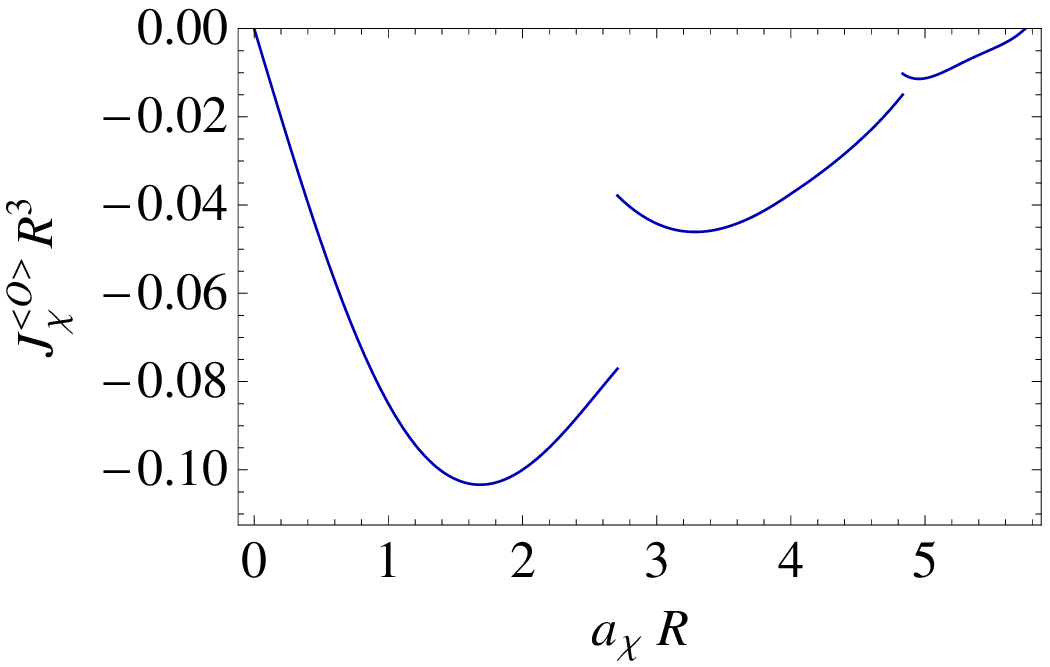}
   \end{tabular}
  \caption{{\footnotesize The current
as a function of $a_{\chi}$ for the solutions of the form (\ref{hol-vortex}). On the left $d=2+1$ with $\mu/ \mu_c=2.1$. On the right $d=3+1$ with $\mu /\mu_c = 1.5$.
We observe a jump when the $m=1$ solution becomes energetically favorable at $a_{\chi} \simeq 4.6/R \, (2.7 /R)$ for $d=2+1 \, (3+1)$. The second 
jump occurs when the $m=2$ solution becomes the ground state.}}
\label{Jvsachi}
\end{figure}

The way to understand these phase diagrams is by following the logic depicted in Section \ref{sec:eft}: by taking into account the Aharonov-Bohm effects one only needs to promote the Ginzbur-Landau parameters $\xi$ and $b$ etc, in the effective action to include a dependence on both $a_\chi$ and $m$ \cite{Montull:2011im}. This leads to several distortions of the phase diagram and of the plots of the order parameter $\OO$ and of the current $J_\chi$ as functions of $a_\chi$, which we detail in Figs. \ref{phase-space-sol}, \ref{Ovsachi} and \ref{Jvsachi}. Let us emphasize that the jumps exhibited by $\OO$ are a manifestation of the non-trivial $m$-dependence of the GL coefficients \cite{Montull:2011im}. Finally, let us point out that while there is a clear dependence of these plots with the dimensionality $d$, this does not seem to significantly change any qualitative features.

At any rate, it is obvious that there is no trace left of the LP periodicity $\Delta a_\chi =1/R$ for the Soliton. The only periodicity that survives in the Soliton case depends on what is the charge of the operator with smallest charge.
This is an additional parameter $e = g_0 /N$, which so far we needed not specify and which we did not fix in the plots, but which is trivially implemented by identifying them periodically with period $\Delta a_\chi =N/R$.\footnote{
Just like in the $\theta$-dependence of Yang-Mills theories \cite{Witten:1998uka}, the fact that the response of the Soliton lacks an explicit periodicity in $a_\chi$ (as seen in Figs. 5-7) is a consequence of the multi-valuedness
of the effective action, that is, of the presence of a tower of excited states. The periodicity in $a_\chi$ is restored once one restricts to be in the ground state for all $a_\chi$.
This is equivalent to the  restriction (and periodic identification) of the plots to the fundamental `domain', $-(\Delta a_\chi)/2 < a_\chi < (\Delta a_\chi)/2$.
}

Hence, we find that the Holographic superconductors fall within the characterization of the flux periodicities given in Sec \ref{sec:eft}. The only peculiarity is that in the BB phase the Aharonov-Bohm effects are suppressed (and so the LP periodicity $\Delta a_\chi =1/R$ is exact for $\NN\to\infty$), whereas in the Soliton phase the Aharonov-Bohm effects are unsuppressed and therefore there is no trace of the LP periodicity.

\section{Response to a Magnetic field} \label{vortices}

Now that we understand the role of the Wilson line $W$ and the winding number $m$ in our model, we are ready to study the impact of magnetic fields along non-compact directions, that extends the phase diagram including a new vortex phase for both, the superconductor and the superfluid cases. We will focus on the solitonic background only, since in the BB background the different topological sectors are degenerate and also turning on a Wilson line produces a very simple modification to the case ($a_\chi=0$) already studied in \cite{Domenech:2010nf} (see also \cite{Albash:2009ix,Montull:2009fe,Albash:2009iq}). 

Our strategy consists on studying first vortex configurations with no Wilson line ($W$) neither winding ($m$) on the compact direction, i.e the simplest case. Later we include them to consider more general possibilities. Also, although the discussion is done in an unspecified dimension $d$, to be as general as possible, we set $d=4$ before any numerical calculations is carried on. All the resulting solution with or without ($W,m$) correspond to physically relevant configurations that describe a vortex within the superconductor phase in the insulator regime of the model \cite{Nishioka:2009zj}. Therefore they represent important new phases, necessary to fully characterize the total phase space diagram of the system. 

To study cylindrical vortex configurations we require our fields to have the form \cite{Montull:2009fe,Albash:2009ix,Albash:2009iq}:
\begin{equation}
\Psi=\psi(r,z)\, e^{in\phi}\ ,\ \ A_0=A_0(r,z)\ , \ \  A_\phi=A_\phi(r,z)\, ,
 \label{ansatz}
\end{equation}
where ($r,\phi$) stand for the radial and angular cylindrical coordinates embedded in the non-compact space. 
The equations of motion for the Ansatz in (\ref{ansatz}) in the corresponding Solitonic background are given by
\begin{eqnarray}
  z^{d - 1} \partial_z \left( \frac{f}{z^{d - 1}} \partial_z \psi \right) +
  \frac{1}{r} \partial_r (r \partial_r \psi) + \left[ A_0^2 - \frac{(A_{\phi}
  - n)^2}{r^2} \right] \psi & = & 0 \hspace{0.25em}, \nonumber\\
  z^{d - 3} \partial_z \left( \frac{f}{z^{d - 3}} \partial_z A_{\phi} \right)
  + r \partial_r \left( \frac{1}{r} \partial_r A_{\phi} \right) - \frac{2
  \hspace{0.25em} (A_{\phi} - n)}{z^2} \hspace{0.25em} \psi^2 & = & 0
  \hspace{0.25em}, \nonumber\\
  z^{d - 3} \partial_z \left( \frac{f \partial_z A_0}{z^{d - 3}} \right) +
  \frac{1}{r} \hspace{0.25em} \partial_r \left( r \partial_r A_0 \right) -
  \frac{2 \hspace{0.25em} A_0}{z^2} \hspace{0.25em} \psi^2 & = & 0
  \hspace{0.25em} .  \label{eom}
\end{eqnarray}
For both the superfluid and the superconductor, we will demand (\ref{bcboth}). We also impose regularity to our solutions. This requires that at $z = z_0$
\begin{eqnarray}
  - \frac{d}{z_0} \partial_z \psi + \frac{1}{r} \partial_r (r \partial_r \psi)
  + \left[ A_0^2 - \frac{(A_{\phi} - n)^2}{r^2} \right] \psi & = & 0
  \hspace{0.25em}, \nonumber\\
  - \frac{d}{z_0} \partial_z A_{\phi} + r \partial_r \left( \frac{1}{r}
  \partial_r A_{\phi} \right) - \frac{2 \hspace{0.25em} (A_{\phi} - n)}{z_0^2}
  \hspace{0.25em} \psi^2 & = & 0 \hspace{0.25em}, \nonumber\\
  - \frac{d}{z_0} \partial_z A_0 + \frac{1}{r} \partial_r \left( r \partial_r
  A_0 \right) - \frac{2 \hspace{0.25em} A_0}{z_0^2} \hspace{0.25em} \psi^2 & =
  & 0 \hspace{0.25em},  \label{bch}
\end{eqnarray}
while at $r = 0$ we must have
\begin{eqnarray}
  \partial_r A_0 & = & 0,\quad A_{\phi} = 0 \hspace{0.25em}, \nonumber\\
  \partial_r \psi & = & 0\quad \text{for}\quad n = 0,\quad \psi = 0 \quad\text{for}\quad n \neq 0
  \hspace{0.25em} .  \label{bcr0}
\end{eqnarray}

In the next sections we will discuss numerical solutions to this set of
equations both in the superfluid and the superconductor case, which we have
found by using  COMSOL  {\cite{COMSOL}}.

\subsection{Holographic superfluid vortices}

{\noindent}For a vortex superfluid configuration $a_{\phi}$ is fixed:
\begin{equation}
  a_{\phi} = A_{\phi} |_{z = 0} = \frac{1}{2} Br^2 \hspace{0.25em},
  \label{bcsf}
\end{equation}
where the constant $B$ represents the external rotation (or, equivalently, the
external magnetic field for a superconductor in a situation in which the
magnetic field can be considered frozen). This corresponds to a Dirichlet boundary condition at $z=0$.

Also we impose the following boundary conditions at $r = r_M$:
\begin{equation}
  \partial_r \psi = 0, \hspace{1em} \partial_r A_0 = 0, \hspace{1em} A_{\phi}
  = \frac{1}{2} B r_M^2 \hspace{0.25em} . \label{BCr=RSF}
\end{equation}
These conditions are consistent with the variational principle which is used
to derive the equations of motion from the action. The first two conditions
represent the physical requirement that, far away from the vortex center, the
solution should reduce to the superconducting/superfluid phase, which is
independent of $r$, while the third one is a simple option compatible with (\ref{bcsf}).

\begin{figure}[t]
  \begin{tabular}{cc}
    {\hspace{-0.1cm}}
    \includegraphics[scale=0.70]{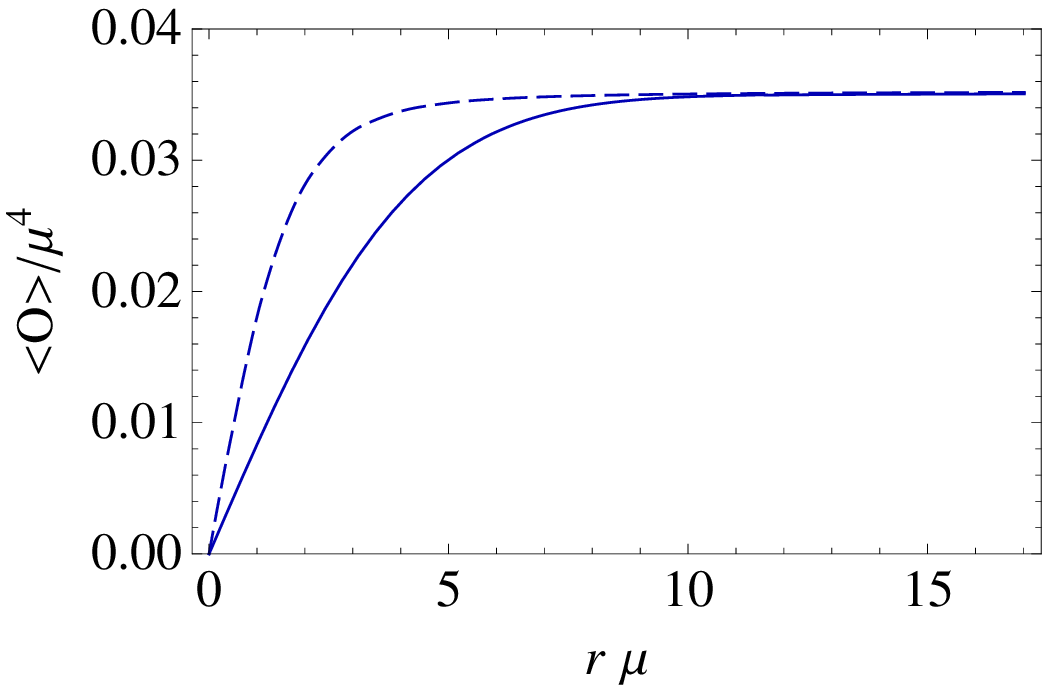}  &
     {\hspace{1.5cm}} \includegraphics[scale=0.70]{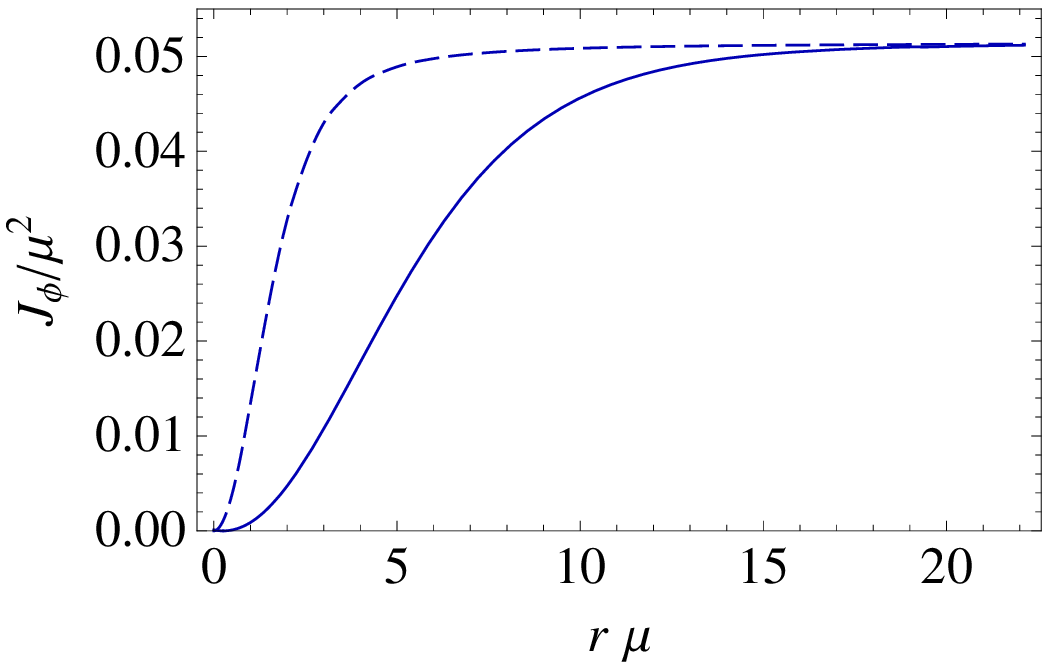}
  \end{tabular}
  \caption{{\footnotesize The modulus of $\langle \mathcal{O} \rangle$ and
  $\langle \hat{J}_{\phi} \rangle$ 
 as
  functions of $r$ from the holographic model in the $n = 1$ superfluid vortex
  solution for $d = 3 + 1$ (solid lines). In this plot we chose $R / R_c =5$ and $B = 0$. The dashed lines are the corresponding profiles in the GL
  model.}}
 \label{O}
\end{figure}

We have solved Eqs. (\ref{eom}) with the boundary conditions in
(\ref{bcboth}), (\ref{bcsf}), (\ref{bch}), (\ref{bcr0}) and (\ref{BCr=RSF}).
In Fig. \ref{O} we give the order parameter and the current as functions of
$r$ for the $n = 1$ vortex solution.

At this point it is interesting to compare our results with the GL
effective theory in (\ref{GL-F}) .  There $\Psi_{\GL}$ is identified with $\sqrt{h_0} \langle
\mathcal{O} \rangle$, with $h_0$ a positive constant. 
For the vortex configuration, the GL current is $J_{\phi}^{\GL} = 2
(n - a_{\phi}) | \Psi_{\GL} |^2$ and we identify it with $\langle
\hat{J}_{\phi} \rangle$. We fit $\xi_{\GL}$ and
$b_{\GL}$ from two predictions of the holographic model:
$B_{c 2}$ and $\langle \hat{J}_{\phi} \rangle$ at large $r$. The quantity
$B_{c 2}$ is determined in the holographic model as the value of $B$ at which
$\langle \mathcal{O} \rangle$ reduces to zero everywhere in space. With this
value we can obtain $\xi_{\GL}$ (see Table \ref{table}). By requiring the current for large $r$ and $B
= 0$ in the holographic model to be equal to the corresponding quantity in the
GL theory, $J_{\phi}^{\GL} (r \rightarrow \infty) = 2 n| \Psi_{\GL} (r
\rightarrow \infty) |^2$, we can then extract $b_{\GL}$. We
observe that the GL curves differ considerably from the holographic ones. In
particular we observe that the radius size of the vortex core is bigger in the
holographic model than in the GL theory, like for the vortices on the
AdS BB geometry {\cite{Domenech:2010nf}}. However, we checked
that, as expected, the GL values for $\langle \mathcal{O} \rangle$ and
$\langle \hat{J}_{\phi} \rangle$ approximate to the corresponding holographic
quantities when we bring the system close to the critical point, $R \simeq
R_c$.

We now turn to the determination of the range of $B$ for which the vortex configurations are energetically favorable. We have seen (Table \ref{table}) that $B_{c1}\rightarrow 0$ as $r_M\rightarrow \infty$
and that the second critical field coincides with the superconductor one, $B_{c 2} = H_{c 2}$,
which we will give\footnote{$H_{c 2}$ for other values of the bulk scalar mass has been computed in \cite{Cai:2011tm}.} in Section \ref{SCvortices}. Since $B_{c 2}$ is
non-vanishing, there exists a finite range of $B$ for which vortex
solutions are energetically favorable. This result was expected because
superfluids can be considered as deep Type II superconductors. 
When $B$ is slightly smaller than $H_{c 2}$ the GL theory can  be applied to predict a triangular lattice of vortices \cite{triangular}, like for the AdS BB \cite{Domenech:2010nf}.
This property has been checked in Ref. \cite{Maeda:2009vf} for the AdS BB in $d=2+1$. We emphasize that it is also true  in arbitrary dimensions and for the AdS soliton background because 
is a consequence of the fact that when $B$ is slightly smaller than $H_{c2}$
the GL theory holds.

\subsection{Holographic superconductor vortices}\label{SCvortices}

\begin{figure}[t]
  \begin{tabular}{cc}
    {\hspace{-0.1cm}} \includegraphics[scale=0.70]{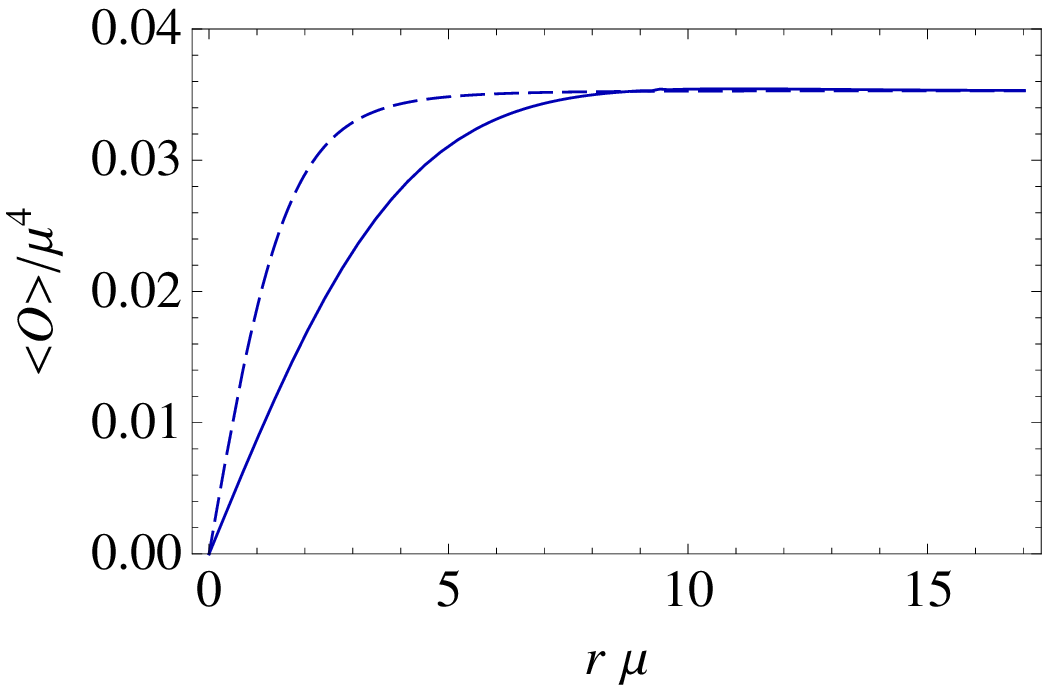}  &
    {\hspace{1.5cm}} \includegraphics[scale=0.70]{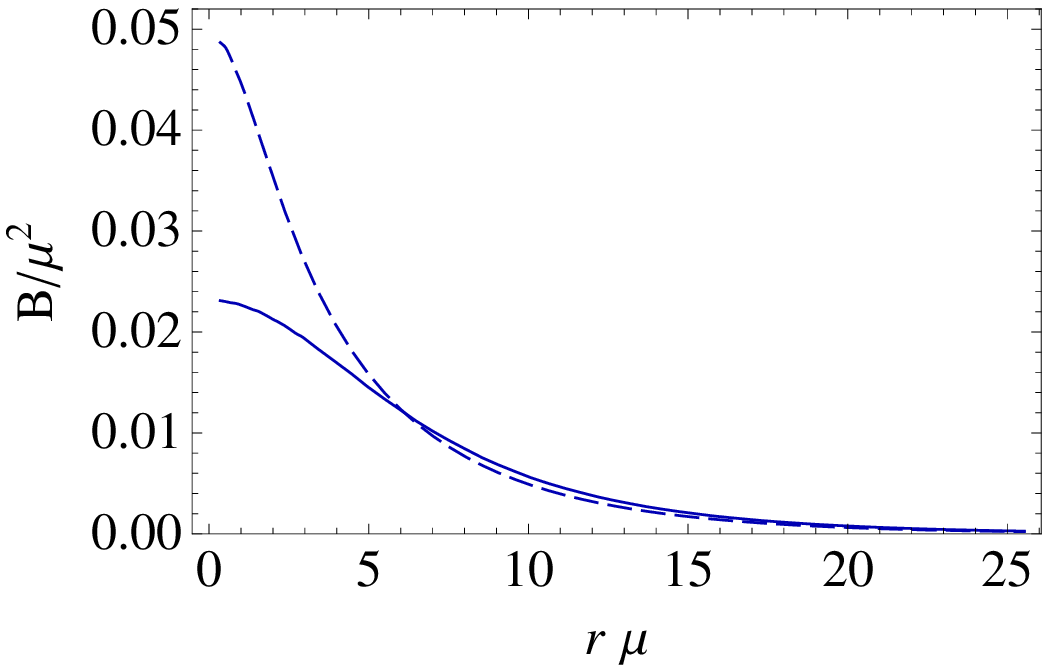}
  \end{tabular} 
  \caption{{\footnotesize The modulus of $\langle \mathcal{O} \rangle$ 
 and $B$ as functions of $r$ from our holographic
  model in the $n = 1$ superconductor vortex solution for $d = 3 + 1$ (solid
  lines). In this plot we chose $R / R_c = 5$ and $e_b$ to satisfy $g_0^{-
  2} (R = R_c) \simeq 1.7 L / g^2$. The dashed lines are the corresponding
  profiles in the GL theory.}}
 \label{OSC}
\end{figure}

{\noindent}To model an Abrikosov vortex we consider stationary configurations
that do not possess a dynamical electric field but only a dynamical magnetic
field. Therefore at $z = 0$ we will impose the boundary condition
Eq.~(\ref{bcboth}) for $A_0$ and Eq.~(\ref{maxwell2}) for $A_i$ that, in polar
coordinates, reads
\begin{equation}
\frac{L^{d-3}}{g^2}  z^{3 - d} \partial_z A_{\phi} \Big{|}_{z = 0} +
  \frac{1}{e_b^2} r \partial_r \left( \frac{1}{r} \partial_r A_{\phi} \right)
  \Big{|}_{z = 0} = 0 \label{newbcsc} \hspace{0.25em},
\end{equation}
where we have taken $J^{\mu}_{ext} = 0$. When $r \rightarrow \infty $ we impose that
\begin{equation}
  \partial_r \psi = 0 \ , \quad \partial_r A_0 = 0 \ , \quad A_{\phi} = n \label{rinftbc}
  \hspace{0.25em} .
\end{equation}

\begin{figure}[t]
  \begin{tabular}{cc}
    {\hspace{-0.1cm}} \includegraphics[scale=0.70]{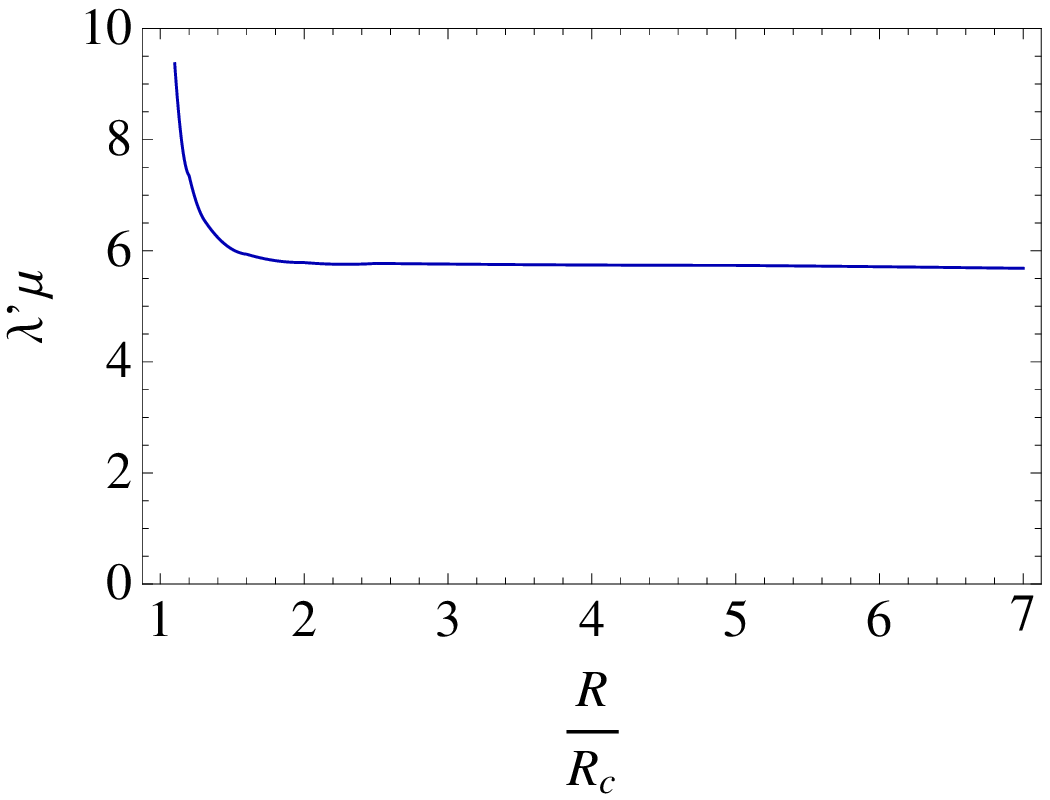}  &
    {\hspace{1.5cm}} \includegraphics[scale=0.70]{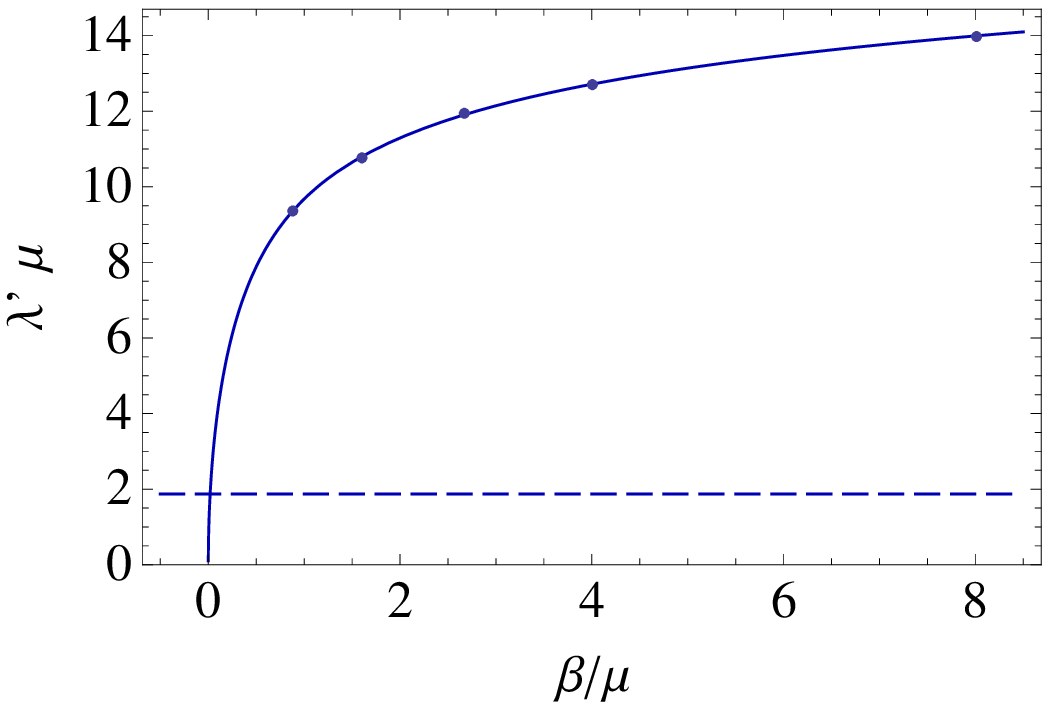}
  \end{tabular}
  \caption{{\footnotesize On the left, we  have $\lambda'$ as a function of $R$ from our holographic
  model for $d = 3 + 1$. We chose $e_b$ to satisfy $g_0^{- 2} (R = R_c) \simeq
 1.7 L / g^2$. On the right, we give $\lambda'$ as a function of $\beta$ for $R/R_c=1.1$; there the dots are  obtained directly from the $a_\phi$-behavior in Table \ref{table}, while the solid line is constructed from Eq. (\ref{lambdaGL}) by computing separately $g_0$ and  $\Psi_{\GL} (r \rightarrow \infty)$. In both plots we have set $d=3+1$. }}
 \label{lambdaprime}
\end{figure}

We have solved Eqs. (\ref{eom}) with the boundary conditions in
(\ref{bcboth}), (\ref{newbcsc}), (\ref{bch}), (\ref{bcr0}) and
(\ref{rinftbc}). In Fig. \ref{OSC} we give the order parameter and the
magnetic field $B (r) = \partial_r A_{\phi} |_{z = 0} / r$ as functions of $r$
for the $n = 1$ vortex solution; by using the profile $B(r)$ one can explicitly see that the total magnetic flux through the vortex line is equal to $2\pi$, namely that $\int dr r B(r)=1$.
We have checked that our solutions
satisfy the $a_\phi$-behavior in Table \ref{table} and we provide $\lambda'$ as a function of $R$
in Fig. \ref{lambdaprime}.   We observe that, as expected, $\lambda' \rightarrow \infty$ as $R \rightarrow R_c$: in this limit the order parameter becomes small and the GL theory can be applied to predict $\lambda' \rightarrow
\infty$. 

 Also, in figure \ref{OSC} we show the corresponding
curves in the GL theory, Eq. (\ref{GL-F}),
where the parameters $\xi_{\GL}$ and $b_{\GL}$ in
the GL potential are fixed as in the superfluid case. The  charge $g_0$ is determined by using the GL relation 
\be \lambda' = 1 / \left( \sqrt{2}g_0 | \Psi_{\GL} (r \rightarrow \infty) |\right) \hspace{0.25em} , \label{lambdaGL} \ee
 and by requiring
$\lambda'$ to be equal to that of the holographic superconductor. Notice that in the  GL case $\lambda'=\lambda$. Again, as in
the superfluid case, we observe that the radius size of the vortex core is
bigger in the holographic model than in the GL theory. As expected, we find
that the differences disappear as $R \rightarrow R_c$.

In the particular case $d=3+1$, our low energy effective theory is defined in 2+1 dimensions.
Hence, as showed in the introduction we have a finite value of $e_{2+1}$  when $R\rightarrow 0$ and the
UV cut-off is removed, according to the presence of an emergent gauge boson.
In the superconducting phase, this property is reflected in the behavior of $\lambda'$ for small $R$ and large $\beta z_0$. For example, in the GL regime , $R\simeq R_c$, the penetration length and the electric charge are related by (\ref{lambdaGL}); 
 the logarithmic running of the coupling constant in the UV corresponds then to a logarithmic dependence of $\lambda'$ on $\beta$.
In Fig. \ref{lambdaprime}, we plot the value of $\lambda'$  as a function of
the renormalization scale, showing precisely this behavior.

It is interesting to know if the superconductors under study are of Type II. The value of $H_{c
2}$ coincides with $B_{c 2}$ of the holographic superfluid, while a formula to compute $H_{c 1}$
is given in Table \ref{table}. $F_n$ is given by
\begin{equation}
  F_n = \frac{T}{V^{d - 3}} S_E + 2 \pi \int drr \frac{1}{2 e_b^2}
  \frac{(\partial_r a_{\phi})^2}{r^2} \hspace{0.25em}, \label{fnsc2}
\end{equation}
where the second term is the contribution of the bare kinetic term in Eq. (\ref{maxwell2}) \cite{Domenech:2010nf}.
\begin{figure}[t]
 \begin{tabular}{cc}
    {\hspace{-0.1cm}}  \includegraphics[scale=0.70]{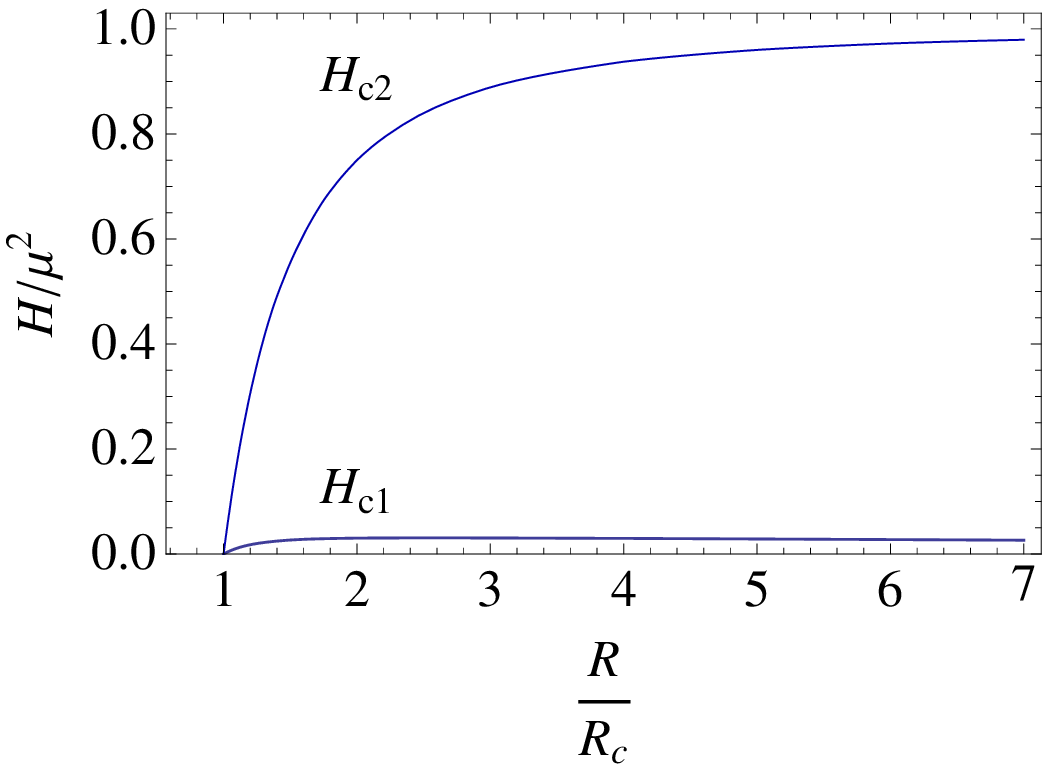} &
     {\hspace{1.5cm}} \includegraphics[scale=0.70]{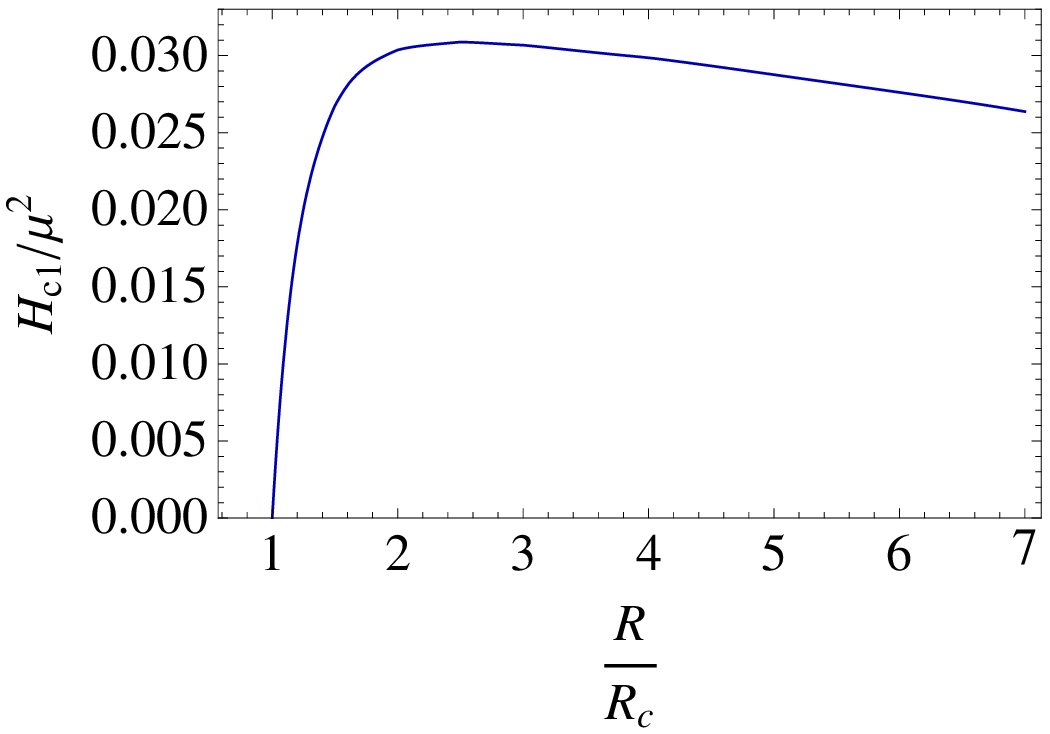}
  \end{tabular}
   \caption{{\footnotesize $H_{c 1}$ and $H_{c 2}$ as functions of $R$ from our
   holographic model for $d = 3 + 1$. We chose $e_b$ to satisfy $g_0^{- 2} (R =
   R_c) \simeq 1.7 L / g^2$.}}  \label{Hc12}
\end{figure}
In Fig. \ref{Hc12} we show $H_{c 1}$ and $H_{c 2}$ as functions
of $R$. Notice that $H_{c 1} \rightarrow 0$ as $R \rightarrow \infty$. This is
due to our normalization of $H$ that makes $H_{c 1}
\propto g_0^2$, which goes to zero as $R \rightarrow \infty$. We can, however,
derive $H_{c 2} / H_{c 1} \rightarrow \infty$ as $R \rightarrow \infty$
independently of such normalization. This is a generic prediction of the
model. Since we have $H_{c 1} < H_{c 2}$ the superconductors under study here
are also, like those introduced in {\cite{Hartnoll:2008vx}}, of Type II. Like for the superfluids, when $H$ is slightly smaller than $H_{c2}$ the GL theory can be applied and predicts
that a Type II superconductor presents a lattice of vortices. Such configuration is therefore the energetically favorable one at $H$ just below $H_{c 2}$ for the Soliton SC,
as well as for the BB SC \cite{Domenech:2010nf}.

\subsection{Vortices and uplifting of the flux period}\label{vortexmn}

Now that we have fully characterized the phase space and the vortex configurations in the simplest scenario on the soliton background, we are ready to study the impact of Wilson lines $W$ and winding number $m$. Recalled what we showed in Section \ref{Wilson}, BB backgrounds are blind to the different topological sectors controlled by $(W,m)$ (due to uniqueness classical theorems). Therefore we are guaranteed to see no differences in solutions that have different $W$ and $m$ but equal $a_\chi-m/R$. On the other hand, due to the uplift of this degeneracy in the solitonic background, we do expect to see differences among physical observables with different $W$ and $m$ even if $a_\chi-m/R$ is the same.  

Therefore, we expect that as we change $(W,m)$, the response of the system to external parameters (like magnetic fields, temperature, size of the compact direction, etc) will change. The simplest way to understand this behavior is perhaps to consider the particular regime in which the effective field theory of the system reduces to the usual GL description. In this case, the topological sectors selected via the external parameters ($W,m$), enter the effective field theory through the coupling constants $\xi_{\GL},b_{\GL}$ of the potential
\be
  V_{\GL} \equiv - \frac{1}{2 \xi_{\GL}(W,m)^2} |
 \Psi_{\GL} |^2 + b_{\GL}(W,m) |
  \Psi_{\GL} |^4\,, \hspace{0.25em} 
\ee
showing us that the vortex solutions will depend on the vacuum sector where they sit on. Nevertheless, we should not forget that the GL approach is valid only near phase transitions while the holographic description is much more general and applies to the whole range of parameters defining the phase space.

To see the uplift of the degeneracy among the different topological sectors in the soliton background, we will study the response of vortex configurations as function of $(W,m)$, comparing the solutions with different winding that would be identified if the degeneracy was not broken (i.e. those with equal $(a_\chi-m/R$). The relevant ansatz for these configurations is given by 
\begin{equation}
  \Psi = \psi (r,z) e^{i (n \phi+m \chi / R)}, \hspace{1em} A_0 = A_0 (r,z),
  \hspace{1em} A_{\phi} = A_{\phi} (r,z), \hspace{1em} A_{\chi} = A_{\chi} (r,z) \hspace{0.25em},
\end{equation}

\begin{figure}[t]
  \begin{tabular}{cc}
    {\hspace{-0.1cm}}
    \includegraphics[scale=0.70]{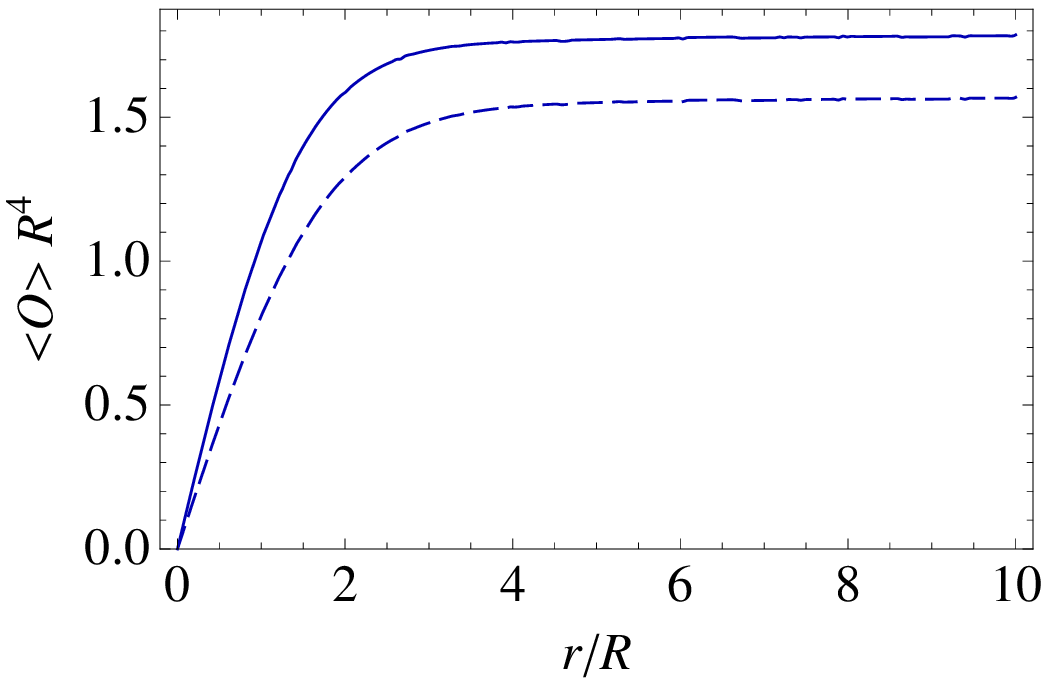}  &
     {\hspace{1.5cm}} \includegraphics[scale=0.70]{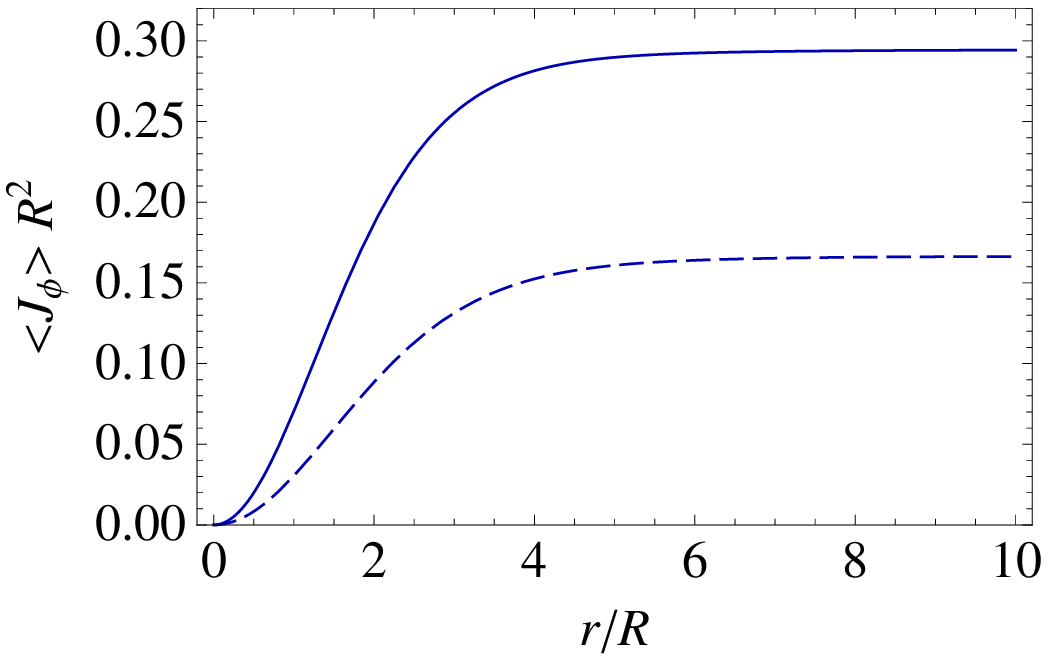}
  \end{tabular}
  \caption{{\footnotesize The modulus of $\langle \mathcal{O} \rangle$ and
  $\langle \hat{J}_{\phi} \rangle$ 
as
  functions of $r$ from the holographic model in the $n = 1$ superfluid vortex
  solution for $d = 3 + 1$. The solid and dashed lines correspond to $(m=0, a_\chi=0)$ and to $(m=1, a_\chi=1/R)$.  In this plot we chose $\mu / \mu_c =1.5$ and $B =0$.}}
\label{mnvortex}
\end{figure}

The equations of motion for the Ansatz are given by
\begin{eqnarray}
  z^{d - 1} \partial_z \left( \frac{f}{z^{d - 1}} \partial_z \psi \right) +
  \frac{1}{r} \partial_r (r \partial_r \psi) + \left[ A_0^2 - \frac{(A_{\phi}
  - n)^2}{r^2}-\frac{(A_\chi-m/R)^2}{f} \right] \psi & = & 0 \hspace{0.25em}, \nonumber\\
  z^{d - 3} \partial_z \left( \frac{f}{z^{d - 3}} \partial_z A_{\phi} \right)
  + r \partial_r \left( \frac{1}{r} \partial_r A_{\phi} \right) - \frac{2
  \hspace{0.25em} (A_{\phi} - n)}{z^2} \hspace{0.25em} \psi^2 & = & 0
  \hspace{0.25em}, \nonumber\\
  z^{d - 3} \partial_z \left( \frac{f \partial_z A_0}{z^{d - 3}} \right) +
  \frac{1}{r} \hspace{0.25em} \partial_r \left( r \partial_r A_0 \right) -
  \frac{2 \hspace{0.25em} A_0}{z^2} \hspace{0.25em} \psi^2 & = & 0
  \hspace{0.25em}, \nonumber\\
  z^{d - 3} \partial_z \left( \frac{\partial_z A_{\chi}}{z^{d - 3}} \right) +
  \frac{1}{r f} \hspace{0.25em} \partial_r \left( r \partial_r A_\chi \right) -
  \frac{2 \hspace{0.25em} (A_{\chi} -m / R)}{z^2 f} \hspace{0.25em}
  \psi^2 & = & 0 \hspace{0.25em}.
\label{mnansatz} 
\end{eqnarray}
We will impose regularity to our solutions. This requires at $z = z_0$
\begin{eqnarray}
    \psi = 0\quad \text{for}\quad m \neq 0\,,\quad - \frac{d}{z_0} \partial_z \psi + \frac{1}{r} \partial_r (r \partial_r \psi)
  + \left[ A_0^2 - \frac{(A_{\phi} - n)^2}{r^2} \right] \psi & = & 0
  \hspace{0.25em}\quad \text{for}\quad m = 0 \hspace{0.25em}, \nonumber\\
  - \frac{d}{z_0} \partial_z A_{\phi} + r \partial_r \left( \frac{1}{r}
  \partial_r A_{\phi} \right) - \frac{2 \hspace{0.25em} (A_{\phi} - n)}{z_0^2}
  \hspace{0.25em} \psi^2 & = & 0 \hspace{0.25em}, \nonumber\\
  - \frac{d}{z_0} \partial_z A_0 + \frac{1}{r} \partial_r \left( r \partial_r
  A_0 \right) - \frac{2 \hspace{0.25em} A_0}{z_0^2} \hspace{0.25em} \psi^2 & =
  & 0 \hspace{0.25em}, \nonumber\\
  A_{\chi} & = & 0 \hspace{0.25em}.
\label{mnbc1} 
\end{eqnarray}
while at $r = 0$ we must have
\begin{eqnarray}
  \partial_r A_0 & = & 0,\quad A_{\phi} = 0,\quad \partial_r A_{\chi} = 0 \hspace{0.25em}, \nonumber\\
  \partial_r \psi & = & 0\quad \text{for}\quad n = 0,\quad \psi = 0 \quad\text{for}\quad n \neq 0
  \hspace{0.25em}.
\label{mnbc2}
\end{eqnarray}

The above set of equations define our cylindrical vortex solutions in different topological sectors labeled by $(W,m)$. As we have pointed out in Section \ref{Wilson}, already at this level, we can explicitly see that there is no gauge transformation that identifies all the different topological sectors, since in the solitonic background the associated gauge transformation is broken due to the form of the boundary conditions.

We have solved equations (\ref{mnansatz}) with boundary conditions (\ref{mnbc1}, \ref{mnbc2}) for the cases ($a_\chi=1/R,m=1$) and $(a_\chi=m=0)$. Notice that these solutions would be ``gauge equivalent'' if the relevant gauge transformation was not broken. In fig. \ref{mnvortex} we show the order parameter $\langle{\cal O}\rangle$ and the current $\langle \hat{J}_{\phi} \rangle$ as a function of $r$, for both cases. As predicted before, the profiles of our observables are indeed sensitive to the topological sector where the vortex is defined, showing a clear signal of the uplifting of the degeneracy among the topological sectors. In other words vortices in the solitonic background can tell, in which sector they sit on while BB vortices can not. 

\section{Conclusions and discussion}\label{conclusions}

In this paper we have studied the magnetic response of holographic superconductors which in the normal phase are insulators.
These materials can be obtained by compactifying a CFT to a cylinder, which is dual  to the AdS Soliton in the gravity side. 
We have studied separately the response to a Wilson line on the circle and to a magnetic field perpendicular to the non-compact directions. 
Continuing the analysis of \cite{Montull:2011im}, we have found that the response to the Wilson line for holographic conductors and holographic insulators 
is dramatically different: at leading order in the large $\NN$ (number of colours) expansion the Aharonov-Bohm effects generated by the Wilson line are suppressed for the conductor and unsuppressed for the insulator. The (un)suppressed Aharonov-Bohm effects leave a clear signature in the superconductivity phase transition in the form of a different periodicity in the cylinder-threading magnetic flux.
%
Regarding the response to the perpendicular magnetic fields, qualitatively there is no great difference between the two types of materials since both of them  respond by creating vortices and are of Type II. Still, we found that for the insulator/superconductor transition the vortices are sensitive to the quantum hair provided by the discrete gauge charge. 
We have also elaborated on the fluid mechanical analogue of the problem, that is, the response to rotation of pairing-based superfluids or supersolids in a cylindrical topology. We concluded similarly that there is an unsuppressed sensitivity to quantum hair (via the Sagnac effect) in supersolids that manifests in a larger periodicity with respect to the angular frequency of rotation.

Our findings agree with the predictions from  condensed matter microscopic theories. Recent literature \cite{Tesanovic,Vakaryuk,Loder,WG,barash} has discussed how the LP period $\Phi^{LP} = 2\pi/g_0$ should 
be uplifted to the fundamental period $\Phi^{fund} = 2\pi/e$ when $R$ is lowered at least down to the zero-temperature
coherence length, $\xi_0 \gtrsim R$. Exactly the  same phenomenon  occurs in the holographic model \cite{Montull:2011im} and we also developed an effective field theory description of it. 
An interesting benefit of  the holographic methods is that,  as a `bonus', one obtains the `prediction' that the Aharonov-Bohm effects are generically bigger for materials with an insulator normal phase than for those with  conductor normal phase.  

One obvious improvement of the present analysis is to include the gravitational back-reaction, which will allow to explore beyond the limit $g\to\infty$ taken here. 
In practice what one should do is to introduce a magnetic field and/or a Wilson line in the set-up of \cite{Horowitz:2010jq}, which studied the gravitational backreaction in the holographic superconductor with a compact space-dimension.
Let us note that Ref. \cite{Brihaye:2011vk} appeared recently taking into account the backreaction in a similar model. However, in Ref. \cite{Brihaye:2011vk} the various fluxoid sectors with $m\neq 0$ are incorrectly treated as gauge-equivalent to $m=0$, so a proper analysis is still necessary. We do not expect that this will modify qualitative features such as the suppression of the  Aharonov-Bohm effects in the BB phase. One can foresee that the phase diagram will be enriched by the superimposition of (de)confinement transitions between the Soliton and BB phases in addition to the superconductivity transitions. 
Another interesting direction to study is how much quantum correction in the bulk change the picture (mapping to  the $1/\NN$ corrections in the CFT), especially in the BB case. Again, one expects only small correction to the present analysis. 
The present paper can also be generalized (i) to extend the holographic Josephson junctions of \cite{Horowitz:2011dz} to superconducting materials with insulating normal phase, which could lead to simple holographic duals to 
so called Superconductor-Insulator-Superconductor (SIS) Junctions, and (ii) to describe p-wave holographic superconductors \cite{pwave}.
Finally, other elaborations such as the inclusion of some fermionic matter could also  be useful to make closer contact with real materials.

Let us end by emphasizing once more the main result of this work, namely  a generic and sharp prediction on superconductivity from holography: 
the magnetic response of superconductors in the form of Aharonov-Bohm effects must be bigger for  for strongly coupled materials which in the normal phase are insulators
than for those which  in the normal phase are conductors. To the best of our knowledge, we ignore if this pattern is already known or if it  occurs in known materials. In any case, it at least offers another means to test the usefulness of the holographic techniques for real-world materials.

\vspace{0.5cm}

{\bf Acknowledgments.} We would like to thank Alex Pomarol, Toni Pineda, Koenraad Schalm, Sergey Si\-bi\-rya\-kov and Jan Zaanen for useful discussions.
This work was supported by the EU ITN ``Unification in the LHC Era", contract PITN-GA-2009-237920 (UNILHC), by MIUR under contract 2006022501, the Spanish Consolider-Ingenio 2010 Programme CPAN (CSD2007-00042) and CICYT under contract FPA 2008-01430.  
O.P. is supported by a Ram\'on y Cajal fellowship (subprograma MICINN-RYC) and thanks the organizers and participants of the 
"Paris Meeting on Holography at Finite Density" (November 16-18, 2011, APC, Paris) for many stimulating discussions. 
M.M. is supported by the ``Universitat Aut\`onoma de Barcelona'' PR-404-01-2/08,
FPU grant AP2007-00420.


\end{document}